\documentclass[conference,compsoc]{IEEEtran}
\IEEEoverridecommandlockouts
\usepackage[utf8]{inputenc}
\usepackage{pmboxdraw}
\usepackage{newunicodechar}
\newunicodechar{►}{\ensuremath{\blacktriangleright}}
\newunicodechar{▼}{\ensuremath{\blacktriangledown}}
\newunicodechar{∩}{\ensuremath{\cap}}
\usepackage{cite}
\usepackage{amsmath,amssymb,amsfonts}
\usepackage{algorithmic}
\usepackage{graphicx}
\usepackage{textcomp}
\usepackage{xcolor}
\usepackage{listings}
\usepackage{tikz}
\usetikzlibrary{positioning,arrows.meta,shapes}
\usepackage{booktabs}
\usepackage{hyperref}

\usepackage{titlesec}
\titlespacing*{\section}{0pt}{8pt plus 2pt minus 2pt}{4pt plus 2pt minus 2pt}
\titlespacing*{\subsection}{0pt}{6pt plus 2pt minus 2pt}{3pt plus 2pt minus 2pt}
\titlespacing*{\subsubsection}{0pt}{4pt plus 2pt minus 2pt}{2pt plus 2pt minus 2pt}

\begin{document}

\title{Authenticated Workflows: A Systems Approach to Protecting Agentic AI}

\author{
\IEEEauthorblockN{Mohan Rajagopalan}
\IEEEauthorblockA{\textit{MACAW Security, Inc.}\\
mohan@macawsecurity.com}
\and
\IEEEauthorblockN{Vinay Rao}
\IEEEauthorblockA{\textit{ROOST.tools}\\
vinay@roost.tools}
}

\maketitle

\begin{abstract}
Agentic AI systems automate enterprise workflows but existing defenses—guardrails, semantic filters—are probabilistic and routinely bypassed. We introduce authenticated workflows, the first complete trust layer for enterprise agentic AI. Security reduces to protecting four fundamental boundaries: prompts, tools, data, and context. We enforce intent (operations satisfy organizational policies) and integrity (operations are cryptographically authentic) at every boundary crossing, combining cryptographic elimination of attack classes with runtime policy enforcement. This delivers deterministic security—operations either carry valid cryptographic proof or are rejected. We introduce MAPL, an AI-native policy language that expresses agentic constraints dynamically as agents evolve and invocation context changes, scaling as O(log M + N) policies versus O(M×N) rules through hierarchical composition with cryptographic attestations for workflow dependencies. We prove practicality through a universal security runtime integrating nine leading frameworks (MCP, A2A, OpenAI, Claude, LangChain, CrewAI, AutoGen, LlamaIndex, Haystack) through thin adapters requiring zero protocol modifications. Formal proofs establish completeness and soundness. Empirical validation shows 100\% recall with zero false positives across 174 test cases, protection against 9 of 10 OWASP Top 10 risks, and complete mitigation of two high impact production CVEs.
\end{abstract}

\section{Introduction}
\label{sec:introduction}

Enterprises are struggling to deploy agentic AI systems in production. While these systems promise to automate complex workflows—managing financial transactions, patient records, and critical infrastructure—they introduce security challenges that existing defenses cannot address. For example, within hours of releasing the OpenAI Atlas browser, researchers demonstrated that malicious instructions embedded in webpage content could trigger the assistant to exfiltrate credentials ~\cite{atlas-cve}. OpenAI's CISO acknowledged that ``prompt injection remains an unsolved problem''~\cite{openai-ciso-atlas}. 

The challenge runs deeper than prompt injection alone. LLMs cannot distinguish instructions from data. Non-deterministic execution paths cannot be predicted or statically analyzed. Multi-turn interactions allow gradual context manipulation where each message appears benign. Existing approaches look for patterns or are probabilistic—both require enumerating attacks, leaving defenders grappling with unknown unknowns.

We eliminate unknown unknowns through a systematic approach: instead of enumerating attacks, we bound the system deterministically. We build on the realization that agentic systems can be abstracted as simple, byzantine, distributed systems -- multiple entities interacting across well defined boundaries. Now security reduces to protecting boundary crossings. Conceptually, our approach is based on satisfying two properties at each boundary crossing : understanding \textbf{intent} ensuring operations satisfy organizational policies, and enforcing \textbf{integrity} to ensure operations are authentic and unmodified. Both are necessary, and neither is sufficient by itself.
This design reduces agentic workflows to a deterministic distributed system where boundaries are guarded by policies and crossings are cryptographically protected. Breaking the system requires breaking cryptography, not crafting clever prompts. Many classes of attack such as identity spoofing, session replay, policy substitution etc are eliminated by design -- they must break cryptographic primitives to succeed, others such as unauthorized data access, privilege escalation, credential exfil etc are blocked by policy at runtime.
Architecturally, \textbf{by-policy enforcement} and \textbf{by-design elimination} deliver complementary, composable defense-in-depth.

We studied 100+ agentic applications and identified key challenges: \textbf{(1) Four simultaneous attack surfaces}—tools, data, prompts, context—where attacks compose across surfaces to achieve objectives impossible through any single surface. \textbf{(2) Heterogeneous frameworks}—Agents are built using protocols (MCP, A2A), LLM interfaces (Claude, OpenAI), orchestrators (LangChain, CrewAI, AutoGen). Heterogeneity is a permanent ecosystem feature. \textbf{(3) Compositional security gaps}—per-framework checks miss violations spanning compositions. \textbf{(4) Developer burden}—requiring security logic at every interaction across heterogeneous frameworks is impractical.

Combined, the attack surface grows faster than tools can address, highlighting a critical gap: a trust layer for AI—security infrastructure positioned between application-layer defenses (don't compose) and infrastructure primitives (lack workflow semantics). 

To address (1) we introduce \textbf{authenticated workflows}—a protocol-level primitive enforcing \textit{intent} and \textit{integrity} at every interaction. We prove the 4 attack surfaces are minimal and complete. Protocol-level positioning addresses heterogeneity (2) and eliminates compositional gaps (3). Thin wrappers across 9 popular frameworks resolve developer burden (4), establishing the foundation for an enterprise AI trust layer.

\textbf{Authenticated workflows} combine cryptography, zero-trust principles, and runtime policy enforcement to ensure each operation is authentic, authorized, and attested. We present a novel design using distributed Policy Enforcement Points (PEPs) embedded at control surfaces, verifying cryptographic proofs independently with sub-millisecond overhead, requiring no centralized infrastructure. Each surface verifies operations independently before execution, creating defense in depth where compromising one surface does not compromise others.

To specify \textit{intent} we introduce MAPL, a new AI-native policy language that lets users express agentic constraints in a dynamic, scalable manner even as agents evolve and invocation context changes. MAPL provides three key capabilities: \textit{composable grammar with inheritance} enabling policies to layer and compose through intersection semantics; \textit{radically reduced specification} scaling from $O(M users \times N resources)$ to $O(\log M + N)$ through hierarchical composition and implicit principal resolution; and \textit{dynamic state via attestations}—signed claims proving operations completed, enabling policies like "export data only after anonymization-completed attestation exists" with cryptographic enforceability. 
These advances overcome limitations of traditional policy systems (OPA, AWS Cedar) in agentic environments.

Our focus is control-flow protection across operational boundaries—tool invocations and data retrievals—complementing Rajagopalan et al. \cite{rajagopalan2025prompts}, which secure data flowing through prompts and context. Together, these deliver a formal trust layer for enterprise AI. We prove practicality through a universal security runtime backing nine frameworks (Section~\ref{sec:enforcement}) via thin adapters (200–500 LOC) requiring no protocol changes—demonstrating framework-agnostic security.

Formally, we prove that the four boundaries are complete and minimal, distributed enforcement is sound, and that policy composition preserves security properties.
Empirically, we validate these properties through comprehensive attack testing and production CVE analysis.

Current agentic defenses—guardrails, prompt filters, semantic analysis (pattern matching, ML models, heuristics)—prioritize precision but deliver low or zero recall. Adversarial prompts, encoding tricks, and novel patterns routinely bypass them via semantic blind spots, while false positives block legitimate operations.
In contrast, our cryptographic enforcement guarantees high precision and high recall: operations either bear valid signatures satisfying policies or are rejected outright. Breaking the system demands shattering cryptographic primitives—not crafting clever patterns. This yields complete violation detection, zero false positives, and deterministic trust via computational hardness.

\noindent\textbf{Contributions.}
We present authenticated workflows—protocol-level primitives enforcing cryptographic verification across four control surfaces (prompts, tools, data, context), proven complete and minimal (Lemma 6). We introduce MAPL, an AI-native policy language providing cryptographic workflow dependencies via attestations and reducing policy specification from $O(M \times N)$ to $O(\log M + N)$ through hierarchical composition (Theorems 1-3). We demonstrate universal deployment across nine heterogeneous frameworks (MCP, A2A, OpenAI, Claude, LangChain, CrewAI, AutoGen, LlamaIndex, Haystack) through thin adapters (200-500 LOC) requiring zero protocol modifications, with distributed Policy Enforcement Points embedded at every boundary providing zero-trust verification with sub-millisecond overhead. Formal proofs establish completeness and soundness (Lemmas 1-7); empirical validation demonstrates deterministic guarantees (100\% recall, 0\% false positives) across 174 test cases covering 9 of 10 OWASP Top 10 risks and complete mitigation of two production CVEs (OpenAI Atlas, GitHub MCP).

\section{The Problem: How Agents Work}
\label{sec:problem}

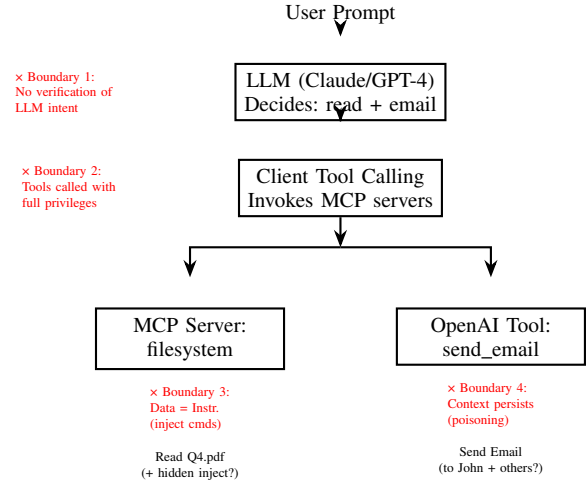
\begin{figure}[t]
\centering
\begin{tikzpicture}[
  node distance=0.5cm,
  >=Stealth,
  box/.style={rectangle, draw, thick, minimum width=2.5cm, minimum height=0.6cm, font=\footnotesize, align=center},
  boundary/.style={font=\tiny, align=left, red}
]

  \node[font=\footnotesize] (userprompt) {User Prompt};
  \draw[->, thick] (userprompt) -- ++(0,-0.3);

  \node[box, below=0.4cm of userprompt] (llm) {LLM (Claude/GPT-4)\\Decides: read + email};

  \node[boundary, left=0.2cm of llm, xshift=-1.3cm] (b1) {× Boundary 1:\\No verification of\\LLM intent};
  \draw[->, thick] (llm) -- ++(0,-0.4);

  \node[box, below=0.5cm of llm] (client) {Client Tool Calling\\Invokes MCP servers};

  \node[boundary, left=0.2cm of client, xshift=-1.3cm] (b2) {× Boundary 2:\\Tools called with\\full privileges};

  \coordinate[below=0.4cm of client] (fork);
  \draw[->, thick] (client) -- (fork);
  \draw[->, thick] (fork) -| ([xshift=-2cm]fork) -- ++(0,-0.4) coordinate (left-end);
  \draw[->, thick] (fork) -| ([xshift=2cm]fork) -- ++(0,-0.4) coordinate (right-end);

  \node[box, below=0.8cm of fork, xshift=-2cm] (fs) {MCP Server:\\filesystem};
  \node[box, below=0.8cm of fork, xshift=2cm] (gmail) {OpenAI Tool:\\send\_email};

  \node[boundary, below=0.1cm of fs] (b3) {× Boundary 3:\\Data = Instr.\\(inject cmds)};

  \node[boundary, below=0.1cm of gmail] (b4) {× Boundary 4:\\Context persists\\(poisoning)};

  \node[font=\tiny, below=0.03cm of b3, align=center] (readop) {Read Q4.pdf\\(+ hidden inject?)};
  \node[font=\tiny, below=0.03cm of b4, align=center] (emailop) {Send Email\\(to John + others?)};

\end{tikzpicture}
\caption{Agentic workflow showing attack cascade across four boundaries.}
\label{fig:problem}
\end{figure}

To understand what makes securing agentic systems hard, consider Figure~\ref{fig:problem}, which shows entities involved when you invoke: ``Read the Q4 financial documents and email a summary to john@company.com.'' The user's prompt flows to the LLM carrying instructions. The LLM reasons and decides to invoke filesystem read and email send operations—MCP tools, OpenAI function calls, or custom implementations. These tool invocations retrieve data from document storage. As the agent executes this multi-turn workflow, it maintains application-level state tracking interaction history.

These interactions cross four fundamental boundaries. \textbf{S1—Prompts} carry instructions into the LLM, directing reasoning and tool selection. \textbf{S2—Tools} execute privileged operations—filesystem access, database queries, API calls, email sending. \textbf{S3—Data} flows from external sources into agent reasoning—document stores, RAG corpora, vector databases, web scraping. \textbf{S4—Context} maintains conversational state across multi-turn interactions. A fundamental challenge: the LLM processes prompts and document contents identically—malicious instructions in Q4.pdf are indistinguishable from user requests. The system operates on implicit trust: context is assumed valid, tools are presumed authorized, data sources treated as benign. These boundaries are framework-agnostic—LangChain chains and CrewAI crews both manifest as prompt→LLM→tool sequences, all crossing the same four surfaces. This universality enables protocol-level security that works uniformly across heterogeneous systems.

  Any surface can serve as an attack entry point with consequences cascading across others. In OpenAI's Atlas attack~\cite{atlas-cve}, malicious instructions embedded in data (S3) flowed into LLM reasoning (S1), generating tool invocations (S2) to exfiltrate credentials, all recorded in context (S4) as normal execution. Different attacks enter through different surfaces. We formalize this through our threat model.


\noindent\textbf{Threat Model.}
\label{sec:threat-model}
We assume a sophisticated adversary and formalize the threat model through adversary capabilities and trust assumptions.

\textbf{Adversary Capabilities} \textbf{A1—Application Control:} Adversaries may control application-layer code. \textbf{A2—Content Injection:} They inject malicious content into data sources, prompts, or context. \textbf{A3—Component Compromise:} They may compromise components and obtain private keys. \textbf{A4—Network Attacks:} They intercept, replay, or modify network traffic. \textbf{A5—Compositional Attacks:} They perform gradual attacks across interactions or span framework boundaries where each operation appears authorized but collectively violates policies.

\textbf{Trust Assumptions} \textbf{L1—Cryptographic Hardness:} Cryptographic primitives provide computational hardness—adversaries cannot forge signatures without private keys. \textbf{L2—Trusted Control Plane:} Control plane services (Agent Registry, Policy Store, Logging, Routing) operate in a minimal trusted computing base with cryptographic integrity guarantees (hash chains for audit logs, Merkle trees for policy store) and administrative access controls. \textbf{L3—Enforcement Integrity:} PEP verification logic executes correctly at framework boundaries. Adversaries with application control (A1) can manipulate business logic but cannot bypass framework APIs or corrupt PEP memory in enterprise deployments where frameworks are immutable and operations route through instrumented APIs. Bypassing PEPs requires compromising framework binaries—equivalent to kernel compromise and out of scope. This parallels OS kernel security. System-level attacks (debuggers, memory corruption) are out of scope—even if such attacks bypass a local PEP, the remote side independently verifies invocations, bounding the adversary to their application scope without gaining new privileges.

\textbf{Security Objectives} Against this threat model, we establish five security objectives: \textbf{O1—Integrity:} Operations are authentic and unmodified. \textbf{O2—Policy Enforcement:} All operations satisfy organizational policies. \textbf{O3—Privilege Non-Escalation:} Composed policies cannot grant broader permissions than individual policies. \textbf{O4—Context Integrity:} Session state is tamper-evident across interactions. \textbf{O5—Accountability:} All operations have non-repudiable audit trails. Section~\ref{sec:formal-analysis} proves authenticated workflows achieve O1-O5 against adversaries with capabilities A1-A5 under assumptions L1-L3.

\textbf{Scope} We focus on protocol-level authorization: cryptographically enforcing that every operation satisfies organizational policy. We provide the mechanism—end users supply the policy.
Application-level safety (e.g., detecting malicious code) requires domain-specific semantics and lies outside protocol scope. We treat all applications as untrusted, bound by system-level policies.

Current approaches are reactive—detecting patterns, filtering inputs, sandboxing execution. Each new attack variant requires new detection rules, creating an enumeration problem for an unbounded attack space. These adversary capabilities (A1-A5) target the four attack surfaces: application control and content injection compromise prompts/tools/data, while compositional attacks exploit multi-framework gaps. \textbf{A1, A3, A4} render observability-based solutions useless. Recent work addressing \textbf{A2} and \textbf{A4} with custom models \cite{camel2023} is probabilistic and cannot guarantee safety deterministically.

\textbf{Our Approach} At the protocol level, agent→tool, agent→LLM, and agent→data operations look identical—a boundary crossing. Our approach converts each of these into an authenticated workflow that is bound by policy (O2), verified for authorization (O1, O3, O4) and attested on execution (O5). In the next section \ref{sec:policy} we describe a policy algebra that enforces O3. By-design mechanisms (cryptographic signatures, hash chains) address component compromise (A3) and network attacks (A4). By-policy mechanisms (runtime verification at boundaries) address application control (A1), content injection (A2), and gradual attacks (A5). Section~\ref{sec:formal-analysis} proves operations either carry valid cryptographic proof satisfying policies, or are blocked before execution—deterministic guarantees, not probabilistic detection.

Revisiting the example attack: Each boundary is protected by independent policies. Even if reasoning is corrupted, restricted documents remain inaccessible. After compromise, the application is restricted to what policy permits. Breaking the system requires simultaneously breaking all enforcement layers—computationally hard.

The problem decomposes into two questions: How do we specify policies that govern boundaries? How do we verify integrity at runtime? We introduce MAPL (Section~\ref{sec:policy}) for expressing intent and authenticated workflows (Section~\ref{sec:authenticated-workflows}) for enforcing integrity, implemented via a universal trust layer (Sections~\ref{sec:trust-layer}--\ref{sec:formal-analysis}).

\section{Specifying Intent: MAPL Policy Language}
\label{sec:policy}

We introduce MAPL, an AI-native policy language to specify intent. Agentic systems exhibit dynamism (agents morph identities, spawn sub-agents, delegate capabilities) and scale (10-100x growth per user) that break traditional policy engines. Three properties challenge existing systems: \textbf{Contextual identity}—agents represent different principals based on runtime context, requiring dynamic principal resolution rather than static WHO bindings. \textbf{Dual perspective}—maintaining integrity requires validating every invocation from both caller intent and resource constraints independently; traditional policies tightly couple caller and resource, where compromising the caller grants resource access. \textbf{Verifiable workflow state}—multi-step workflows need sequential enforcement, since application-reported state is untrusted, we need some form of attestation to verify operation A completed before invoking operation B.

  MAPL addresses these through three design choices: (1) Principals inferred from authenticated context at runtime—enabling dynamic identity without policy updates; (2)
  Caller and resource policies expressed independently, composing via intersection—achieving defense in depth without coordination; (3) Policies reference
  cryptographically verified attestations—enabling provable workflow dependencies. The combination creates a complete, provable, compositional model for expressing
  intent.

\noindent\textbf{Policy Grammar.} MAPL (MAPL Agentic Policy Language) policies follow a structured grammar enabling machine verification and automated composition:

\begin{lstlisting}[basicstyle=\ttfamily\footnotesize]
Policy {
  policy_id: <unique_identifier>,
  extends: <parent_policy_id>,
  resources: [<resource_patterns>],
  denied_resources: [<denial_patterns>],
  constraints: {
    parameters: {<resource>: {<param>: [<patterns>]}},
    denied_parameters: {<resource>: {<param>: [<patterns>]}},
    attestations: [<required_attestation_names>]
  }
}
\end{lstlisting}

A MAPL policy includes: \textit{policy\_id} a unique identifier for composition chains and cryptographic binding, \textit{extends} the parent policy reference for hierarchical composition, \textit{resources} allowed operation patterns; wildcards \texttt{*}/\texttt{**} match single/recursive levels, \textit{denied\_resources} explicit blocks overriding allowances, and \textit{constraints} (parameters , denied parameters, patterns and attestation requirements). Appendix~\ref{appendix:policy-examples} demonstrates concrete policies.

\noindent\textbf{Hierarchical Composition.} The \textbf{extends} field enables policies to inherit from parent policies through intersection semantics. This allows organizational policies to layer (base → department → team) where each child policy refines its parent by adding restrictions, never relaxing constraints. When a policy extends another, the effective permissions are the intersection of parent and child—maintaining monotonic restriction as the composition algebra formalizes below.

\textbf{Expressiveness}. The minimal grammar suffices through four mechanisms: (1) Hierarchical resources with wildcards express unbounded namespaces; (2) Positive and negative specifications express arbitrary boolean combinations; (3) Parameter constraints express any decidable predicate on arguments; (4) Attestations enable sequential constraints through cryptographic state. This covers organizational hierarchies, dual-perspective defense, workflow dependencies, exception patterns, and parameter controls.

\textbf{Composition Algebra.} We formalize how policies compose at runtime and prove the system satisfies key security properties.

\textbf{Policy Structure and Operations.}

A policy $P = (R, D, C)$ consists of:
\begin{itemize}
\item $R$: Allowed resource patterns
\item $D$: Denied resource patterns
\item $C$: Operational constraints (parameters, attestations)
\end{itemize}

\textbf{Policy Intersection} $P_1 \cap P_2 = (R', D', C')$ where:
\begin{itemize}
\item $R' = R_1 \cap R_2$ (allow only if both policies permit)
\item $D' = D_1 \cup D_2$ (deny if either policy forbids)
\item $C' = \text{MostRestrictive}(C_1, C_2)$ (apply tightest constraint)
\end{itemize}

MostRestrictive$(C_1, C_2)$ takes minimum values for numeric limits, intersection for allowed patterns, and union for required attestations and denied patterns.

Policy Enforcement Points compose policies from organizational hierarchy and dual perspectives at runtime. This ensures operations can only add restrictions, never relax them. If a policy is absent, it contributes no constraints (most permissive interpretation violating no stated constraint).

\textbf{Formal Security Properties}

{\small
\noindent Define the permission function:
\[
\pi(P) = \{(r, op) \mid r \notin \text{Match}(D) \land r \in \text{Match}(R) \land op \text{ satisfies } C\}
\]

\noindent\fbox{\begin{minipage}{0.96\columnwidth}
\textbf{Theorem 1 (Monotonic Restriction):} For composition $P_0 \cap \ldots \cap P_n$:
\vspace{-0.5em}
\[
\forall i, j : (i < j) \Rightarrow \pi(P_0 \cap \ldots \cap P_j) \subseteq \pi(P_0 \cap \ldots \cap P_i)
\]
\vspace{-0.3em}

\textit{Proof Sketch:} By construction, $P_{i+1} = P_i \cap P_{\text{next}}$. Resource intersection: $R_{i+1} \subseteq R_i$. Denial union: $D_{i+1} \supseteq D_i$. Therefore $\pi(P_{i+1}) \subseteq \pi(P_i)$. By induction, $\pi(P_j) \subseteq \pi(P_i)$ for $i < j$. $\square$
\end{minipage}}

\vspace{0.15em}

\noindent\fbox{\begin{minipage}{0.96\columnwidth}
\textbf{Theorem 2 (Transitive Denial):} If resource $r$ is denied by any policy, it remains denied:
\vspace{-0.3em}
\[
\exists i : r \in \text{Match}(D_i) \Rightarrow r \in \text{Match}(D_{\text{eff}})
\]
\vspace{-0.3em}

\textit{Proof Sketch:} $D_{\text{eff}} = D_0 \cup D_1 \cup \ldots \cup D_n$. If $r \in D_i$ for any $i$, then $r \in D_{\text{eff}}$ by set union. $\square$
\end{minipage}}

\vspace{0.15em}

\noindent\fbox{\begin{minipage}{0.96\columnwidth}
\textbf{Theorem 3 (No Privilege Escalation):} If base policy $P_0$ denies $r$, no composition grants access:
\vspace{-0.5em}
\[
(r \in \text{Match}(D_0)) \Rightarrow r \notin \text{Allowed}(P_{\text{eff}})
\]
\vspace{-0.3em}

\textit{Proof:} By Theorem 2, if $r \in \text{Match}(D_0)$, then $r \in \text{Match}(D_{\text{eff}})$. By definition of $\pi$, denied resources cannot be allowed. $\square$
\end{minipage}}
}
\textbf{Security Implications}
These theorems provide formal guarantees: \textbf{Theorem 1} prevents privilege expansion—adding policies can only narrow permissions; \textbf{Theorem 2} ensures any layer's denial is absolute; \textbf{Theorem 3} prevents escalation where combining policies grants denied permissions.

Together, these achieve policy enforcement (O2) and privilege non-escalation (O3) from Section~\ref{sec:threat-model}. Even when adversaries compromise components and obtain their keys (A3), the compromised component can only perform operations within its policy-permitted scope—monotonic restriction prevents privilege expansion, and transitive denial ensures root denials propagate through all derivations. These properties hold mathematically—independent of attacker sophistication or LLM behavior. Breaking them requires breaking cryptographic primitives (digital signatures, hash functions), not manipulation.
\textbf{No arbitrary overrides}. MAPL explicitly disallows overrides—they break provability where Theorems 1-3 do not hold. Instead, administrators create temporary groups with time-bounded validity (analogous to Unix groups/permissions). These compose via intersection, maintaining formal guarantees while supporting operational flexibility. Appendix~\ref{appendix:policy-examples} demonstrates time-bounded emergency access.

\noindent\textbf{Enabling Scale: From O(MxN) rules to O(log M + N) policies}
\label{sec:practical-policy}
Traditional policy engines require O(M×N) explicit rules for M principals and N resources. 
Managing rules that scale combinatorially is impractical; MAPL provides a practical solution as enterprises scale agents across thousands of users.
MAPL's hierarchical composition and dynamic principal binding eliminates this explosion by mirroring organizational structure -- with a branching factor of 8-16, departments scale logarithmically: $D \approx \log_{8-16}(M)$—exponentially smaller than M, teams inherit upwards absorbing into the hierarchical tree. Resources grow with functionality, independent of user count. Thus MAPL requires $\log(M) + N$ policies -- a significant, practical reduction.


Together, MAPL's compositional algebra addresses all three agentic requirements: contextual identity through runtime principal resolution, dual perspective through independent policy composition, and verifiable workflow state through cryptographic attestations.

\section{Authenticated Workflows}
\label{sec:authenticated-workflows}

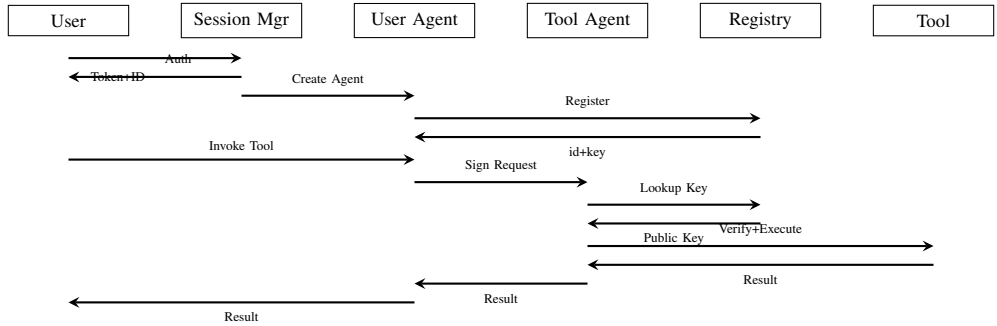
\begin{figure*}[t]
\centering
\begin{tikzpicture}[
  node distance=0.2cm,
  entity/.style={rectangle, draw, minimum width=1.6cm, minimum height=0.4cm, font=\scriptsize, align=center},
  arrow/.style={->, thick, >=stealth},
  label/.style={font=\tiny}
]

\node[entity] (user) at (0,0) {User};
\node[entity] (session) at (2.3,0) {Session Mgr};
\node[entity] (useragent) at (4.6,0) {User Agent};
\node[entity] (toolagent) at (6.9,0) {Tool Agent};
\node[entity] (registry) at (9.2,0) {Registry};
\node[entity] (tool) at (11.5,0) {Tool};

\draw[arrow] (0,-0.5) -- node[label, right] {Auth} (2.3,-0.5);
\draw[arrow] (2.3,-0.75) -- node[label, left] {Token+ID} (0,-0.75);

\draw[arrow] (2.3,-1.0) -- node[label, above] {Create Agent} (4.6,-1.0);

\draw[arrow] (4.6,-1.3) -- node[label, above] {Register} (9.2,-1.3);
\draw[arrow] (9.2,-1.55) -- node[label, below] {id+key} (4.6,-1.55);

\draw[arrow] (0,-1.85) -- node[label, above] {Invoke Tool} (4.6,-1.85);
\draw[arrow] (4.6,-2.15) -- node[label, above] {Sign Request} (6.9,-2.15);
\draw[arrow] (6.9,-2.45) -- node[label, above] {Lookup Key} (9.2,-2.45);
\draw[arrow] (9.2,-2.7) -- node[label, below] {Public Key} (6.9,-2.7);

\draw[arrow] (6.9,-3.0) -- node[label, above] {Verify+Execute} (11.5,-3.0);
\draw[arrow] (11.5,-3.25) -- node[label, below] {Result} (6.9,-3.25);
\draw[arrow] (6.9,-3.5) -- node[label, below] {Result} (4.6,-3.5);
\draw[arrow] (4.6,-3.75) -- node[label, below] {Result} (0,-3.75);

\end{tikzpicture}
\caption{Registration and invocation flow showing authentication, entity registration, signed invocation, and bidirectional verification.}
\label{fig:registration-invocation}
\end{figure*}

We now introduce authenticated workflows—a protocol-level mechanism that realizes deterministic verification at every agentic boundary crossing.
Against adversaries with capabilities A1-A5 (Section~\ref{sec:threat-model}), authenticated workflows provide four \textbf{cryptographic guarantees} that together achieve security objectives O1-O5:

\textbf{Authenticity} (achieves O1: Integrity): Every accepted invocation is cryptographically linked to the principal that initiated it and the operation being performed. Attackers cannot forge invocations without possessing the principal's private key, binding both WHO (principal) and WHAT (operation).

\textbf{Policy Binding} (achieves O2-O3: Policy Enforcement, Privilege Non-Escalation): The policy governing an operation is cryptographically bound to the invocation, preventing attackers from substituting policies without invalidating signatures.

\textbf{Tamper Evidence} (achieves O4: Context Integrity): Any modification to invocations, context state, or policy bindings is cryptographically detectable through hash chains and sequence numbers.

\textbf{Non-Repudiation} (achieves O5: Accountability): Digital signatures create undeniable proof linking principals to their operations, enabling forensic analysis and compliance verification with tamper-evident audit logs.

As with authenticated system calls \cite{rajagopalan2006,erlingsson2004}, the core idea is conceptually simple: augment every inter-entity invocation with a policy and cryptographic signature that ensures integrity.
{\small
\begin{verbatim}
Invocation = (Args, Policy, MAC)
MAC = Sign(K, Args || Policy)
\end{verbatim}
}
where the Message Authentication Code(MAC) signature cryptographically binds the arguments and policy identifier together using a secret or private key $K$. On the receiving end, verification proceeds in three steps: validate the signature to ensure authenticity and integrity, retrieve and verify the policy binding to prevent substitution, then evaluate whether the operation satisfies policy rules.

\subsection{Design Principles}

Realizing authenticated workflows requires four principles that compose to deliver the cryptographic guarantees:

\textbf{Zero-Trust Identity}: Every entity—agents, tools, data sources, LLMs—possesses unique cryptographic identity (keypair). Even entities in the same process verify each other independently. No implicit trust propagates. We bridge enterprise identity with agent-scale verification: External principals authenticate via enterprise IAM; the runtime propagates these identities as tamper-proof session context. Internally, lightweight ephemeral signatures identify agents, addressing dynamic identity challenges (Section~\ref{sec:policy}).

\textbf{Boundary Verification}: Every inter-entity communication—prompt→LLM, agent→tool, tool→data—requires independent cryptographic verification. Operations carry signatures and policy identifiers; receivers verify signatures using registered public keys, retrieve policies, and evaluate constraints. Verification is transport-agnostic—identical across MCP, LangChain, OpenAI APIs, or custom protocols.

\textbf{Policy Enforcement Points}: PEPs provide independent verification at every boundary—entities cannot verify their own operations. Deployed in-process (linked library) or as sidecars, PEPs verify invocations before execution, providing horizontal scalability, zero-trust verification, tamper resistance, and enforcement independence. This relies on L3 (enforcement integrity).

\textbf{Trustworthy Authenticated Context}: Runtime provides tamper-evident session state such as from~\cite{rajagopalan2025prompts} that ensures integrity including binding operations to session scope (context IDs, session IDs), preventing replay (sequence numbers), detecting tampering (hash-chains), and enabling workflow dependencies (attestations proving prerequisite operations completed).

\noindent\textbf{The Protocol.}
These principles compose into a four-phase protocol that every boundary crossing is an authenticated workflow. Figure~\ref{fig:registration-invocation} shows the complete flow.

\textbf{Registration}: Each entity registers, receiving (agent\_id, keypair). Registration happens at finest granularity—individual tools within agents receive independent (tool\_id, keypair), minimizing blast radius.

\textbf{Invocation}: Caller constructs signed invocation binding operation, arguments, policy, and session context.

\begin{figure}[t]
\centering
\begin{tikzpicture}[
  node distance=0.4cm,
  box/.style={rectangle, draw, minimum width=4.5cm, minimum height=0.5cm, font=\scriptsize, align=center},
  decision/.style={rectangle, draw, minimum width=2cm, minimum height=0.45cm, font=\tiny, align=center},
  arrow/.style={->, thick, >=stealth}
]

\node[box] (receive) at (0,0) {Invocation Received at PEP};

\node[box] (stage1) at (0,-0.8) {Stage 1: Signature Verification\\{\tiny Authenticity + Integrity}};
\node[decision] (reject1) at (3.5,-0.8) {Invalid?\\REJECT};

\node[box] (stage2) at (0,-1.6) {Stage 2: Policy Binding Verification\\{\tiny Prevent Substitution}};
\node[decision] (reject2) at (3.5,-1.6) {Mismatch?\\REJECT};

\node[box, minimum height=0.7cm] (stage3) at (0,-2.6) {Stage 3: Policy Evaluation\\{\tiny Resources, Parameters, Attestations}};
\node[decision] (reject3) at (3.5,-2.6) {Violation?\\REJECT};

\node[box, minimum height=0.7cm] (pre) at (0,-3.6) {Pre-Invocation Custom Verifiers\\{\tiny PII redaction, Prompt safety}};

\node[box] (execute) at (0,-4.5) {Execute Operation};

\node[box, minimum height=0.7cm] (post) at (0,-5.5) {Post-Invocation Custom Verifiers\\{\tiny Secrets removal, Output validation}};

\node[box] (sign) at (0,-6.3) {Sign Result \& Return};

\draw[arrow] (receive) -- (stage1);
\draw[arrow] (stage1) -- node[right, font=\tiny] {Valid} (stage2);
\draw[arrow] (stage2) -- node[right, font=\tiny] {Valid} (stage3);
\draw[arrow] (stage3) -- node[right, font=\tiny] {Pass} (pre);
\draw[arrow] (pre) -- (execute);
\draw[arrow] (execute) -- (post);
\draw[arrow] (post) -- (sign);

\draw[arrow] (stage1.east) -- (reject1.west);
\draw[arrow] (stage2.east) -- (reject2.west);
\draw[arrow] (stage3.east) -- (reject3.west);

\end{tikzpicture}
\caption{PEP verification flow with three-stage cryptographic verification and optional custom verifiers.}
\label{fig:verification-flow}
\end{figure}
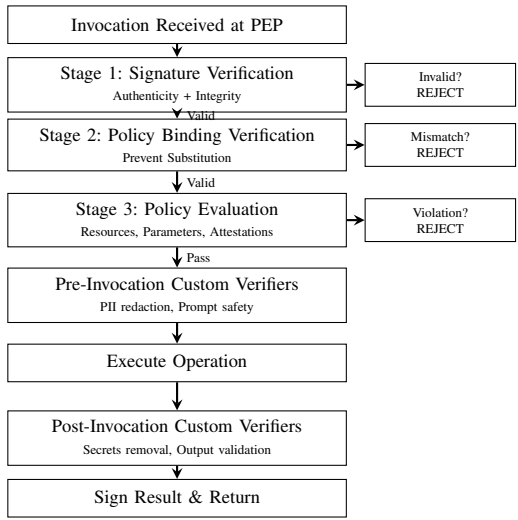

\textbf{Verification}: The receiving PEP independently verifies through three stages: (1) signature verification (authenticity + integrity), (2) policy binding verification (prevent substitution), (3) policy evaluation (resource permissions, parameter constraints, attestations). Figure~\ref{fig:verification-flow} details verification.

\textbf{Attestation}: Upon completion, the service signs the result. Context is updated with cryptographically signed attestation proving the operation completed, maintaining Authenticated Context for downstream operations.

Services sign results; callers verify service signatures before accepting results—realizing bidirectional authentication analogous to mutual TLS. This stateless protocol ensures each hop is independently verified without implicit trust propagating through multi-step workflows.

\subsection{Verification Flow}

Figure~\ref{fig:verification-flow} shows the verification pipeline—the heart of our integrity claim. The pipeline is always enforced and cannot be escaped: every invocation undergoes cryptographic verification before execution, with receivers independently verifying callers using public keys retrieved from the registry.

\textbf{Policy Construction}: Stage 3 constructs the effective policy independently at the receiver through dual-perspective enforcement. The PEP retrieves organizational policies (company, business unit, team) and resource-specific constraints, computing $P_{\text{eff}} = \text{Intent} \cap \text{Resource}$ (Section~\ref{sec:policy}). The caller's claimed intent is independently verified against resource constraints—callers cannot influence policy evaluation. The intersection algebra (Theorems 1-3) guarantees monotonic restriction: composed policies only narrow permissions, never broaden them.

\textbf{Custom Verifiers}: MAPL's declarative constraints handle resource permissions and parameters, but some security checks require imperative logic on invocation context. Custom verifiers extend policy evaluation with PII detection, SQL injection prevention, prompt safety via LLM classifiers (LLMs as judges), rate limiting, and geolocation checks. Pre-invocation verifiers execute after policy evaluation; post-invocation verifiers process results. Operating on cryptographically verified context, verifiers are trusted code checked in by system administrators, not application developers. Once declared in the verification flow, the PEP enforces their execution—they cannot be bypassed. This provides domain-specific extensibility (HIPAA, GDPR, SOC 2) while maintaining integrity guarantees.

Four production verifiers demonstrate this architecture: (1) MemoryIntegrityVerifier prevents memory poisoning (OWASP LLM01) through rate limiting, protected field enforcement, and cryptographic integrity checks; (2) WorkflowIntegrityVerifier prevents workflow hijacking (OWASP LLM04) by enforcing prerequisite steps and maintaining attestation chains; (3) ToolAuthorizationVerifier prevents tool misuse (OWASP LLM06) via RBAC and dangerous pattern detection (e.g., file\_write→command\_execute); (4) StorageIntegrityVerifier prevents data exfiltration (OWASP LLM02) through path traversal prevention and encryption enforcement. Appendix~\ref{appendix:verifiers} provides implementation details.

This separation is crucial: the core verification pipeline (Stages 1-3: signature verification, policy binding, policy evaluation) provides deterministic guarantees—100\% recall with zero false positives through cryptographic enforcement of MAPL's declarative constraints (Lemmas 1, 4). Custom verifiers (Stages 4-5) are optional administrative extensions that may use heuristics (ML-based PII detection, LLM safety classifiers) and could introduce false positives if misconfigured, but cannot weaken core guarantees—they can only add restrictions, never remove them. Verifiers execute only after cryptographic verification passes, operating on authenticated invocations and tamper-evident state, not unbounded application input. Even if a verifier fails or is compromised, Stages 1-3 still enforce cryptographic integrity and policy constraints. Bypassing verifiers requires corrupting trusted administrator code (L3), not manipulating application logic (A1).


\noindent\textbf{Security Guarantees.}Against adversaries A1-A5 (Section~\ref{sec:threat-model}), the protocol mechanisms compose into defense-in-depth through four independent cryptographic layers: \textbf{Integrity} (signatures establish authenticity, defeating network attacks A4); \textbf{Authorization} (policy intersection enforces intent—operations must satisfy composed policies even with valid signatures, preventing privilege escalation under application control A1); \textbf{Provenance} (hash chains and sequence numbers make context tamper-evident, defeating content injection A2 and multi-turn manipulation A5); \textbf{Compliance} (custom verifiers add domain-specific validation on cryptographically verified context, providing non-repudiation). Together these achieve security objectives O1-O5.

The protocol provides defense-in-depth: each layer operates independently at each boundary (Lemma 7, Section~\ref{sec:formal-analysis}). Compromising signature verification doesn't bypass policy evaluation; tampering with context doesn't defeat tool PEPs; bypassing one verifier doesn't invalidate cryptographic checks. Section~\ref{sec:formal-analysis} formalizes these independence guarantees.


\begin{table*}[t!]
\centering
\scriptsize
\begin{tabular}{llp{11cm}}
\toprule
\textbf{Framework} & \textbf{Layer} & \textbf{Mapping to Authenticated Workflow Primitives} \\
\midrule
MCP & Protocol & Server→Agent, Tools→Tools (w/PEP), Resources→Tools (policy-differentiated), Prompts→AuthenticatedPrompt \\
A2A & Protocol & Agents→Agents (cryptographic identity), Delegation→Signed tokens (scope constraints), Attestations→Attestations (workflow dependencies) \\
OpenAI & LLM API & Endpoint→Agent, Operations (generate/complete/embed)→Tools (w/PEP), Two deployment variants (two-sided/one-sided wrapping) \\
Claude & LLM API & Endpoint→Agent, Operations (generate/complete/embed)→Tools (w/PEP), Two deployment variants (two-sided/one-sided wrapping) \\
LangChain & Orchestration & Agents→Agents, Tools→Tools (w/PEP), Chains→Policy intersection, Memory→AuthenticatedContext (hash chains prevent poisoning) \\
CrewAI & Orchestration & Crew members→Agents, Roles→Attestations (cryptographically bound), Tasks→Signed invocations (role-based authorization) \\
AutoGen & Orchestration & Conversation agents→Agents, Code execution→Tool (w/PEP), Policies→Pluggable verifiers (AST analysis, allowlists, sandbox constraints) \\
LlamaIndex & Orchestration & RAG pipeline→Agent, Query/Retrieve operations→Tools (w/PEP enforcing document access control to prevent prompt injection via retrieval) \\
Haystack & Orchestration & Document pipeline→Agent, Pipeline nodes→Tools (w/PEP), Sequential execution→Policy intersection (nodes cannot relax constraints) \\
\bottomrule
\end{tabular}
\caption{Framework mapping to authenticated workflow primitives.}
\label{tab:framework-mapping}
\end{table*}

\section{A Universal Security Abstraction}
\label{sec:enforcement}

Authenticated workflows provide a universal security abstraction for agentic AI—
We validate this claim by securing nine frameworks spanning three architectural layers: protocol standards (MCP, A2A), LLM APIs (OpenAI, Claude), and orchestration frameworks (LangChain, CrewAI, AutoGen, LlamaIndex, Haystack). 
Regardless of framework architecture, every agentic interaction maps completely to a verifiable boundary crossing governed by our four-phase protocol: Registration → Invocation → Verification → Attestation.

\subsection{Framework Integration}
Table~\ref{tab:framework-mapping} shows how nine frameworks map to authenticated workflow primitives. We highlight representative examples at each architectural layer:

\textbf{Protocol Layer: MCP}.
MCP servers expose tools, resources and prompts.
Resources capture static data-sets intended to give LLMs additional context to work with, while MCP tools are active functions with side effects, and parameters.
Servers map to agents within our system, prompts map to Authenticated Prompts \cite{rajagopalan2025prompts}, and resources and tools are abstracted as verified tools guarded by different types of policies  --  each with a unique independent identity and key pairs.

\noindent\begin{minipage}[t]{\columnwidth}
{\scriptsize
\begin{verbatim}
MCP Server → Agent with Registered Tools
┌────────────┬──────────────┬────────────────┐
│   Tools    │  Resources   │    Prompts     │
│            │              │                │
│ execute()  │  read()      │  getPrompt()   │
│ params     │  watch()     │  arguments     │
│            │              │  template      │
│            │              │                │
│     ↓      │      ↓       │       ↓        │
│  Tool      │  Tool        │ Authenticated  │
│  (w/PEP)   │  (w/PEP)     │ Prompt         │
│            │ (policy-diff)│ (signed)       │
└────────────┴──────────────┴────────────────┘
\end{verbatim}}
\end{minipage}

\textbf{Protocol Layer: A2A}. A2A—the agent-to-agent protocol—maps directly to authenticated workflows. Agents register with cryptographic identity. Delegation becomes signed tokens with scope constraints following MAPL's intersection algebra—delegates cannot grant broader permissions than received. Attestations enable workflow dependencies through cryptographic proof, ensuring operations complete in required order.

\textbf{LLM Interfaces: OpenAI and Claude}. OpenAI and Claude follow identical patterns. API endpoints register as agents. LLM inference operations (generate, complete, embed) register as tools with embedded PEPs. Two deployment variants provide flexibility: \textit{two-sided wrapping} treats both the LLM API and client application as agents—the API-side PEP verifies incoming requests satisfy provider policies, while the client-side PEP verifies function calls satisfy application policies; \textit{one-sided wrapping} wraps only client-side tools, treating the LLM API as passthrough, simplifying deployment when the LLM provider is trusted for prompt security and the focus is protecting client side tool calling. This addresses the challenge that LLMs are untrusted for authorization decisions—verification must happen at function execution boundaries (S2) where PEPs independently verify signatures and evaluate policies.

\textbf{Orchestration Layer: LangChain}. Orchestration frameworks coordinate multi-step workflows across tools and LLMs, exposing multiple control points: chain composition logic, agent reasoning, memory operations, inter-agent communication. Securing only individual tools would leave composition and state management unprotected.

Our design addresses security at four layers. Agents register with cryptographic identity. Tools register independently with embedded PEPs—each tool invocation undergoes signature verification and policy evaluation. Chains—sequential compositions of operations—enforce policy composition through intersection (Section~\ref{sec:policy}); if a tool policy denies operations, composing that tool into a chain cannot relax the restriction—effective policy becomes more restrictive through intersection, never more permissive. Memory—persistent state across interactions—enforces context integrity through authenticated context~\cite{rajagopalan2025prompts}, preventing context poisoning (S4).

\noindent\begin{minipage}[t]{\columnwidth}
{\scriptsize
\begin{verbatim}
LangChain Orchestrator (Agent with Tools)
┌──────────┬─────────-┬──────────┬────────-──┐
│  Agents  │  Tools   │  Chains  │  Memory   │
│          │          │          │           │
│Coordinate│Individual│Sequential│Persistent │
│workflow  │operations│compose   │state      │
│          │          │          │           │
│   ↓      │    ↓     │    ↓     │    ↓      │
│  Agent   │  Tool    │  Policy  │Authentic- │
│          │ (w/PEP)  │  ∩       │ated       │
│          │          │          │Context    │
│          │          │          │           │
│Each agent│Each tool │Effective │Context    │
│has own   │verifies  │policy =  │integrity  │
│identity  │signed    │∩ of all  │verified   │
│+ policy  │invoc.    │tool      │every      │
│          │before    │policies  │transition │
│          │execution │          │           │
└──────────┴─────────-┴──────────┴──────────-┘
\end{verbatim}}
\end{minipage}

Other orchestration frameworks (CrewAI, AutoGen, LlamaIndex, Haystack) follow similar patterns—roles map to attestations for cryptographic role binding, code execution uses pluggable verifiers for AST analysis, RAG pipelines enforce document access control as detailed in Appendix ~\ref{appendix:framework-details} 

\subsection{Protocol-Level Universality}
At the protocol level, all nine frameworks collapse to a uniform abstraction: agents invoke operations through signed invocations; PEPs verify signatures and policies; operations execute or are rejected. This uniformity enables consistent security across heterogeneous compositions—LangChain orchestrating OpenAI invoking MCP tools—with independent verification at each boundary.

When workflows span frameworks, compromising one framework's verification does not bypass others. Each boundary's PEP independently verifies: (1) signature validity (authenticity); (2) identity (preventing impersonation); (3) policy evaluation of resource permissions, parameter constraints, and attestations (authorization). This compositional approach—securing boundaries rather than frameworks—enables uniform protection across heterogeneous systems. Section~\ref{sec:formal-analysis} formalizes these guarantees through Lemmas 6-7.

\section{The Agentic AI Trust Layer}
\label{sec:trust-layer}

In this section we describe an Agentic AI Trust Layer, which includes authenticated workflows and authenticated prompts and context as described in~\cite{rajagopalan2025prompts}. All nine frameworks map onto the trust layer using thin framework adapters. The trust layer comprises a control plane providing core services (identity, policy, logging, routing), manages integrations with enterprise infrastructure (IAM, PKI, storage, etc), a verification gateway (sidecar PEP) and client libraries that embed PEPs within agentic applications. Due to space we focus our discussion on key design elements instead of an exhaustive overview.

\begin{figure}[t]
\begin{tikzpicture}[
  agent/.style={rectangle, draw, minimum width=1.1cm, minimum height=0.4cm, font=\tiny, align=center},
  integration/.style={rectangle, draw, minimum width=7.5cm, minimum height=0.4cm, font=\tiny, align=center},
  service/.style={rectangle, minimum width=1.65cm, minimum height=0.5cm, font=\tiny, align=center},
  adapter/.style={rectangle, draw, minimum width=5.5cm, minimum height=0.4cm, font=\tiny, align=center}
]

\node[agent] (agent1) at (-2.4,0) {Agent\\(LLM)};
\node[agent] (agent2) at (-1.2,0) {Agent\\(MCP)};
\node[agent] (agent3) at (0,0) {Agent\\(LangChain)};
\node[agent] (agent4) at (1.2,0) {Agent\\(Tool)};
\node[agent] (agent5) at (2.4,0) {Agent\\(SDK)};

\node[adapter] (adapters) at (-0.6,-1) {Framework Adapters (OpenAI, MCP, etc.)};

\node[draw, thick, rectangle, minimum width=7.5cm, minimum height=0.8cm] (controlplane-box) at (0,-2.3) {};

\node[font=\tiny\bfseries] at (0,-1.95) {Control Plane};

\node[service] (registry) at (-2.25,-2.35) {Agent\\Registry};
\node[service] (policy) at (-0.7,-2.35) {Policy\\Store};
\node[service] (logging) at (0.85,-2.35) {Logging};
\node[service] (routing) at (2.4,-2.35) {Routing};

\node[integration] (integrations) at (0,-3.4) {Enterprise Integrations (IAM, PKI, SIEM, Observability)};

\draw[->, thick, dotted] (agent1.south) -- (agent1.south |- adapters.north);
\draw[->, thick, dotted] (agent2.south) -- (agent2.south |- adapters.north);
\draw[->, thick, dotted] (agent3.south) -- (agent3.south |- adapters.north);
\draw[->, thick, dotted] (agent4.south) -- (agent4.south |- adapters.north);

\draw[->, thick, dotted] ([xshift=-1.2cm]adapters.south) -- ([xshift=-1.2cm]adapters.south |- controlplane-box.north);
\draw[->, thick, dotted] (adapters.south) -- (adapters.south |- controlplane-box.north);
\draw[->, thick, dotted] ([xshift=1.2cm]adapters.south) -- ([xshift=1.2cm]adapters.south |- controlplane-box.north);

\draw[->, thick, dotted] (agent5.south) -- (agent5.south |- controlplane-box.north);

\draw[->, thick, dotted] (controlplane-box.south) -- (integrations.north);

\end{tikzpicture}
\caption{Agentic AI Trust Layer.}
\label{fig:runtime-architecture}
\end{figure}
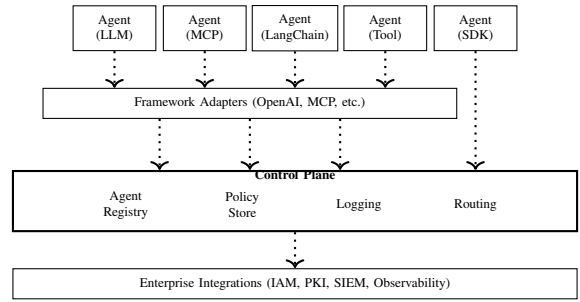

\subsection{Architecture}

Figure~\ref{fig:runtime-architecture} shows the trust layer architecture with agents interfacing with the control plane.

\textbf{Framework Adapters}. Each framework integrates through thin adapters that translate framework-native operations into authenticated workflows. Adapters implement the four-phase protocol—registration, invocation signing, verification, and attestation—while presenting the framework's native API. Integration is transparent; developers use standard framework APIs (MCP servers, OpenAI clients, LangChain agents) while cryptographic enforcement happens automatically.

\textbf{Control Plane Services}. The control plane provides four core services:
\begin{itemize}
\item \textit{Agent Registry}: Stores agent identifiers and public keys, and tool identifiers and public keys within agents, preserving cryptographic lineage. Enables instant credential revocation. PEPs retrieve public keys during signature verification.

\item \textit{Policy Store}: Tamper-proof policy store where PEPs retrieve and compose hierarchical policies (e.g., company, business unit, team) through intersection.

\item \textit{Routing}: Routes signed invocations between agents across network boundaries, enabling callees to access callers' public keys for verification.

\item \textit{Logging}: Records all security events with cryptographic lineage through tamper-evident hash chains. Integrates with enterprise SIEM platforms. Blocked operations generate structured logs indicating rejection reason (signature invalid, policy violation, missing attestation, verifier failure) with sufficient detail for debugging while avoiding policy leakage. Audit logs enable root-cause analysis and compliance reporting.
\end{itemize}

\textbf{Distributed Enforcement}. These services run in highly available configurations, integrating with enterprise infrastructure (IAM, PKI, SIEM, observability platforms). The control plane is multi-tenant—organizations deploy shared infrastructure while maintaining cryptographic isolation between tenants. The architecture realizes zero-trust distributed verification—each PEP independently verifies invocations using public keys from the registry and policies from the store. 
PEPs are embedded in-process by default (linked library) or deployed as sidecars, providing horizontal scalability. 
With embedded PEPs, no centralized bottleneck exists; verification happens locally at each boundary with sub-millisecond overhead. 
Administrators can instantly revoke credentials, update policies, or isolate agents through the control plane. Due to space constraints, full architectural details are out of scope; we refer interested readers to our technical report for complete implementation details.

\subsection{Design Decisions}

Realizing authenticated workflows and our trust layer in production exposed several practical challenges that needed an opinionated design—we discuss six of the most interesting design decisions addressing these challenges.

\textbf{Identity Bridging}. Enterprise identities need bridging across frameworks—each has its own authentication model (OAuth, API keys, SAML). Instead of solving this per-framework, we bridge at the protocol level. When users authenticate, the trust layer issues ephemeral keypairs bound to their sessions and propagates cryptographically-verified principals across all frameworks. Signed invocations carry user identity through delegation chains; each PEP independently verifies the originating principal from the Agent Registry. This centralizes IAM integration—frameworks don't individually integrate with corporate identity systems; the trust layer handles authentication once. When OAuth-authenticated Alice invokes API-key-authenticated Claude which calls MCP-authenticated tools, logs show "Alice via Claude accessed file" preserving audit trails. Protocol-level bridging provides a simpler programming model with consistent principal propagation, eliminating per-framework integration complexity.

\textbf{Session Management}. Many frameworks lack native session concepts (MCP, LangChain are stateless), yet cross-turn correlation is essential for security. Instead of modifying each framework to add sessions, we track at the protocol level using context identifiers and sequence numbers transparent to frameworks. The trust layer maintains authenticated context regardless of framework support—frameworks without sessions receive invisible protocol-level tracking; frameworks with sessions map directly. Multi-tenant architecture ensures cryptographically isolated contexts per session, preventing cross-session contamination. This enables unified audit trails across multi-framework workflows without requiring protocol modifications to individual frameworks.

\textbf{Verifiable Delegation}. Multi-hop delegation creates privilege escalation risks—compromised intermediate agents could claim broader permissions than granted. Instead of trust-based delegation, we enforce cryptographic scope narrowing. Signed delegation tokens contain delegator identity, delegate identity, scope constraints, and expiration. The trust layer enforces scope narrowing via intersection: delegated scope = what delegator grants $\cap$ what delegate possesses, extending MAPL's algebra (Section~\ref{sec:policy}). Compromised agents cannot claim broader permissions than granted; downstream services cryptographically verify delegation validity. This provides provable least privilege across heterogeneous frameworks.

\textbf{Forward Secrecy}. Long-lived API keys create operational and security risks—keys rarely rotate, and a single leak compromises all sessions. Instead of relying on operational key rotation policies, we use ephemeral keys by design. Each session receives unique ephemeral keypairs that expire automatically. Key rotation happens naturally via session lifecycle without infrastructure modifications, providing session-level forward secrecy. Compromising one session's key doesn't affect other sessions, limiting blast radius without operational overhead.

\textbf{Service Accounts at Scale}. Agentic systems spawn entities dynamically at runtime—orchestrators spawn agents, agents spawn tools, workflows create sub-workflows. Traditional IAM's manual provisioning doesn't scale to runtime dynamics. Instead of manual provisioning, we automatically provision cryptographic identities during registration (Section~\ref{sec:authenticated-workflows}) at tool-level granularity—assigning (tool\_id, keypair) and recording in the Agent Registry. No manual provisioning, no credential distribution, no IAM bottleneck. Each tool maintains independent cryptographic accountability, minimizing blast radius. This enables dynamic agent deployment at enterprise scale.

\textbf{Bidirectional Accountability}. Unidirectional authentication creates accountability gaps in agentic environments—callers cannot prove services performed operations correctly; services cannot prove callers requested operations. Instead of one-way signatures, we use dual signatures for mutual non-repudiation. Callers sign requests; services sign results. Each operation produces cryptographic proof from both parties, providing compliance-grade audit trails—services cannot repudiate actions; callers cannot repudiate requests. Combined with identity bridging and tamper-evident logging, this delivers complete accountability across heterogeneous authentication boundaries.

\section{Formal Analysis}
\label{sec:formal-analysis}

We prove authenticated workflows achieve O1-O5 against adversaries A1-A5 under assumptions L1-L3 via seven lemmas organized around intent and integrity. Integrity properties (Lemmas 1, 2, 3) ensure operations are authentic and tamper-evident. Intent properties (Lemmas 4, 5) ensure operations satisfy policies and compose safely. Completeness properties (Lemmas 6, 7) ensure all operations are verified and verification is independent.

\noindent\textbf{Main Security Theorem.} \textbf{Theorem (Authenticated Workflows Security)}: Under trust assumptions L1-L3, authenticated workflows achieve integrity (O1), policy enforcement (O2), privilege non-escalation (O3), context integrity (O4), and accountability (O5) against adversaries with capabilities A1-A5.

\noindent\textbf{Integrity Properties.} These lemmas ensure operations are authentic, tamper-evident, and non-repudiable—addressing adversaries with component compromise (A3) and network attack (A4) capabilities.

\textbf{Lemma 1 (Authenticity)}: All accepted invocations have valid cryptographic signatures binding the principal to the operation, arguments, policy, and context state.

\textit{Proof}: Policy Enforcement Points verify signatures using principals' public keys retrieved from the Agent Registry (Section~\ref{sec:enforcement}). Under cryptographic hardness assumption L1, forging signatures without the private key is computationally infeasible—even adversaries with application control (A1), network access (A4), or compromised components (A3) cannot produce valid signatures for other principals' identities. This achieves integrity (O1): operations are cryptographically authentic. $\square$

\textbf{Lemma 2 (Tamper Evidence)}: Modifications to context state or audit logs are cryptographically detectable, and replay attacks are prevented.

\textit{Proof}: Context state is maintained through hash chains linking sequential states: $h_i = \text{Hash}(h_{i-1} \parallel \text{state}_i \parallel \text{sequence}_i)$. Each invocation includes a monotonically increasing sequence number bound in the signature. Modifying any state requires finding hash collisions; replaying old invocations is detected because PEPs reject sequence numbers less than or equal to previously processed values. Under cryptographic hardness (L1), finding collisions is computationally infeasible. Multi-turn manipulation attacks (A5) attempting gradual state poisoning or replayed operations are detected because tampering breaks the hash chain or violates sequence monotonicity. Audit logs use the same hash chain mechanism. This achieves context integrity (O4). $\square$

\textbf{Lemma 3 (Non-Repudiation)}: Principals cannot deny performing operations recorded in audit logs with valid signatures.

\textit{Proof}: Only the principal possessing the private key can generate valid signatures (Lemma 1). Audit logs maintain tamper-evident records (Lemma 2). Therefore, signed operations in audit logs constitute cryptographic proof of principal actions. This achieves accountability (O5). $\square$

\noindent\textbf{Intent Properties.} These lemmas ensure operations satisfy policies and compose safely—addressing adversaries with application control (A1), content injection (A2), and gradual attack (A5) capabilities.

\textbf{Lemma 4 (Policy Enforcement)}: All executed operations satisfy the effective policy composed from organizational hierarchy and resource constraints, with the policy identifier cryptographically bound in the signature.

\textit{Proof}: Each invocation includes a policy identifier in the signed payload. The PEP retrieves policies and computes the effective policy $P_{\text{eff}}$ through intersection of all applicable policies—organizational (base, department, team) and resource-specific (Section~\ref{sec:policy}). Adversaries attempting policy substitution attacks (capabilities A1, A4) fail because modifying the policy identifier invalidates the signature (Lemma 1). Policy intersection ensures monotonic restriction (Section~\ref{sec:policy}, Theorems 1-3)—composed policies can only become more restrictive, never more permissive. This achieves policy enforcement (O2) and privilege non-escalation (O3). $\square$

\textbf{Lemma 5 (Composition Safety)}: When workflows span multiple frameworks, security properties are preserved across framework boundaries.

\textit{Proof}: Consider operation O1 on framework F1 invoking operation O2 on framework F2. Both invocations carry signatures binding principals and policies (Lemma 1). Framework F2's PEP independently verifies O2's signature and computes effective policy as the intersection of F1's policy and F2's policy (Lemma 4). Framework F1 cannot bypass F2's verification because: (a) F2's PEP verifies signatures independently (Lemma 7, proven below), and (b) policy intersection ensures O2 must satisfy both F1 and F2 constraints. Composition across N frameworks creates N independent verification points with composed policy $P_{\text{eff}} = P_1 \cap P_2 \cap \ldots \cap P_N$, achieving monotonic restriction. Section~\ref{sec:enforcement} demonstrates this across nine framework combinations. $\square$

\noindent\textbf{Completeness Properties.} These lemmas ensure all operations are verified and verification is independent across control surfaces.

\textbf{Lemma 6 (Surface Completeness and Minimality)}: The four control surfaces \{S1, S2, S3, S4\} are complete (all resource access operations cross at least one surface) and minimal (each surface is necessary).

\textit{Proof by Enumeration and Necessity}:

\textit{Completeness:} We enumerate all resource access operations and show each crosses at least one surface: (1) \textit{Computational Resources}—LLM inference crosses S1 (prompts carry instructions); tool execution crosses S2 (privileged operations). (2) \textit{Data Resources}—RAG retrieval, database queries, web scraping cross S3 (external data flows into reasoning)—exploited by content injection attacks (A2). (3) \textit{State Resources}—Session state, workflow context, memory cross S4 (persistent state across turns)—targeted by multi-turn manipulation (A5). (4) \textit{Cross-Agent Communication}—Delegation crosses S1 (instruction propagation) or S2 (invocations) plus S4 (context inheritance). Operations not accessing external resources (pure computation) require no protection—they cannot exfiltrate data or violate policies.

\textit{Minimality:} Each surface is necessary—removing any surface leaves attacks unprotected: Remove S1 (Prompts): indirect prompt injection via S3 data→LLM bypasses verification. Remove S2 (Tools): unauthorized tool execution despite prompt verification. Remove S3 (Data): poisoned RAG data→tool invocations despite S1/S2 verification. Remove S4 (Context): session hijacking and context poisoning across multi-turn workflows. Therefore, \{S1, S2, S3, S4\} is complete and minimal. Section~\ref{sec:enforcement} empirically validates this across nine frameworks spanning three architectural layers. $\square$

\textbf{Lemma 7 (PEP Independence)}: Compromising one Policy Enforcement Point does not weaken verification at other PEPs.

\textit{Proof}: Each PEP verifies operations independently using only: (1) cryptographic primitives for signature and hash chain validation (L1), (2) public keys and policies retrieved from the trusted control plane (L2), and (3) its own verification logic (L3). PEPs share no runtime state and perform no coordination. An adversary compromising PEP at one surface (capability A3) can only bypass verification at that surface—other surfaces perform independent verification using their own PEP instances. Multi-surface attacks require compromising multiple independent PEPs, each protected by L3. This realizes defense in depth: blast radius is limited to the compromised PEP's surface. $\square$

\noindent\textbf{Main Theorem Proof.} \textit{Proof of Main Theorem}: By Lemma 6, all resource access crosses at least one control surface. Each surface has an embedded PEP enforcing independent verification (Lemmas 1-4, 7). Multi-framework workflows compose verification across framework boundaries (Lemma 5). Verification enforces intent via policy evaluation (Lemma 4) and integrity via signatures (Lemma 1) and hash chains (Lemma 2). Attestations provide workflow dependencies through signed, hash-chained claims.

Every resource access operation has one of two outcomes: (1) \textit{Authorized Execution}—operation satisfies all cryptographic checks (valid signature, tamper-free context, valid sequence number) and policy constraints (allowed resource, parameters within bounds, required attestations present), executes, and is logged with non-repudiable audit trail (Lemma 3); (2) \textit{Blocked}—operation fails at least one check (invalid signature, policy violation, tampered context, sequence violation, or missing attestation), rejected before execution with violation logged.

\textit{Concrete Example}: In the Q4 attack (Section~\ref{sec:problem}), poisoned data instructs the LLM to exfiltrate credentials. When the LLM generates a filesystem read for \texttt{credentials.db}, the tool's PEP evaluates policy—which denies paths matching \texttt{*credential*}—blocking the operation despite valid signature from the authenticated LLM. Policy enforcement prevents unauthorized access regardless of signature validity. Even if multiple boundaries are compromised, each surface enforces independent verification (Lemma 7), limiting blast radius.

Therefore, under L1-L3, adversaries with capabilities A1-A5 cannot perform unauthorized resource access without detection, achieving O1-O5. $\square$

\noindent\textbf{Practical Implications.} These formal properties enable the operational solutions in Section~\ref{sec:trust-layer}: Lemma 1 enables principal propagation, Lemmas 4+5 enable verifiable delegation, Lemma 2 enables session-level forward secrecy, and Lemma 3 enables bidirectional accountability.

\section{Attack-Defense Validation}
\label{sec:attack-defense}

\begin{table*}[t]
\centering
\caption{OWASP Top 10 for LLM Applications 2025 Coverage}
\label{tab:owasp-coverage}
\scriptsize
\begin{tabular}{|p{3cm}|p{4cm}|p{6.5cm}|p{1.5cm}|}
\hline
\textbf{OWASP Risk} & \textbf{Our Attack Tests} & \textbf{Defense Mechanism} & \textbf{Section} \\
\hline
LLM01: Prompt Injection & Prompt Injection, Atlas & By-policy: Policy blocks unauthorized operations + MemoryIntegrityVerifier prevents goal manipulation & §4, §7 \\
\hline
LLM02: Info Disclosure & Token Hijack, Data Exfil, Side Channel, Inference & By-policy: Resource denial + StorageIntegrityVerifier detects exfiltration patterns & §4, §5 \\
\hline
LLM03: Supply Chain & Rogue Tool, Supply Chain & By-design + By-policy: Code signing + tool registry with approval policies & §5, §7 \\
\hline
LLM04: Data Poisoning & Context tamper & By-design: Hash chains + WorkflowIntegrityVerifier enforces step sequences & §4, §7 \\
\hline
LLM05: Output Handling & (Cross-cutting) & By-policy: Post-invocation verifiers sanitize outputs & §5 \\
\hline
LLM06: Excessive Agency & Keys to Kingdom, Confused Deputy & By-policy: Policy composition + ToolAuthorizationVerifier enforces RBAC with pattern detection & §3, §7 \\
\hline
LLM07: Prompt Leakage & Prompt variants & By-design: Cryptographic segmentation (AuthPrompt) & §4 \\
\hline
LLM08: Vector Weak. & (RAG scenarios) & By-policy: Attestations verify sources; policies restrict access & §3, §5 \\
\hline
LLM09: Misinform. & (Tangential) & Pluggable verifiers; not primary security focus & §5 \\
\hline
LLM10: Unbounded Cons. & Denial of Service & By-policy: Rate limiting + resource quotas & §5 \\
\hline
\end{tabular}

\raggedright
\scriptsize{Coverage: 9 of 10 OWASP risks. LLM04 partial (runtime poisoning prevented; training data out of scope). LLM09 tangential (quality vs. security).}
\end{table*}

Section~\ref{sec:formal-analysis} proved authenticated workflows achieve O1-O5 against adversaries A1-A5 under assumptions L1-L3. We validate these formal guarantees through OWASP Top 10 coverage (9 of 10 risks), systematic attack testing (11 patterns, 174 test cases, 100\% recall, 0\% false positives), and production CVE analysis demonstrating protection where baseline systems failed.

\noindent\textbf{OWASP Top 10 for LLM Applications 2025 Coverage.} Table~\ref{tab:owasp-coverage} maps our defenses to the OWASP Top 10 for LLM Applications 2025~\cite{owasp-llm-2025}, demonstrating systematic coverage with explicit defense mechanisms and formal guarantees.

These defenses address all adversary capabilities (Section~\ref{sec:threat-model}): by-policy mechanisms block A1, A2, A5 through runtime enforcement (Lemmas 4, 5); by-design mechanisms eliminate A3, A4 through cryptographic hardness (Lemmas 1, 2, 3). Together, they achieve O1-O5. This validates the dual framework: neither mechanism alone suffices; both together provide complete protection.

\noindent\textbf{Systematic Attack Validation.} Beyond OWASP's broad risk categories, we analyze attack mechanisms—how attacks manifest in workflow composition. We identify 8 mechanism categories spanning prompt manipulation, tool chaining, credential theft, data exfiltration, malware/code execution, resource exhaustion, multi-agent attacks, and policy violations. Detailed taxonomy appears in Appendix C. Table~\ref{tab:attack-patterns} shows 11 attack patterns providing representative coverage.

\begin{table*}[t]
\centering
\caption{Validated Attack Patterns and Defense Classification}
\label{tab:attack-patterns}
\scriptsize
\begin{tabular}{|p{3.5cm}|p{3.5cm}|p{2.5cm}|p{3cm}|}
\hline
\textbf{Attack Pattern} & \textbf{Category} & \textbf{Defense Class} & \textbf{OWASP} \\
\hline
Prompt Injection & Prompt Manipulation & By-Policy & LLM01 \\
\hline
Keys to Kingdom & Tool Chaining & By-Policy & LLM06 \\
\hline
Confused Deputy & Tool Chaining & By-Policy & LLM06 \\
\hline
Token Hijacking & Credential Theft & By-Policy & LLM02, LLM06 \\
\hline
Session Fixation & Credential Theft & By-Design & Beyond OWASP \\
\hline
Data Exfiltration & Data Exfiltration & By-Policy & LLM02 \\
\hline
Side Channel & Data Exfiltration & By-Policy & LLM02 \\
\hline
Inference Attack & Data Exfiltration & By-Policy & LLM02 \\
\hline
Rogue Tool & Malware/Code Exec & By-Policy & LLM03 \\
\hline
Supply Chain & Malware/Code Exec & By-Policy & LLM03 \\
\hline
Denial of Service & Resource Exhaustion & By-Policy & LLM10 \\
\hline
\multicolumn{4}{|l|}{\small\textit{Total: 11 patterns, 18 explicit variants, 174 test cases including framework permutations}} \\
\hline
\end{tabular}

\raggedright
\scriptsize{Coverage: 6 of 8 mechanism categories explicitly implemented; Multi-Agent and Policy Violation attacks addressed through compositional mechanisms (see text).}
\end{table*}

Each pattern includes multiple test scenarios: Data Exfiltration tests bulk export, multi-resource access, PII extraction, unauthorized access; Inference Attack tests attribute, membership, reconstruction attacks. Our 174 test cases span framework combinations (OpenAI, Anthropic, LangChain, etc) and configuration variants.

Table~\ref{tab:attack-patterns} shows defense classification. By-design mechanisms realize integrity properties (Lemmas 1, 2, 3), rendering attacks cryptographically impossible: identity spoofing, session replay, policy substitution, context tampering, audit manipulation, attestation forgery. By-policy mechanisms realize intent properties (Lemmas 4, 5), blocking violations through runtime verification: prompt injection, privilege escalation, credential exfiltration, data harvesting, supply chain, resource exhaustion, cross-agent attacks. Complex attacks like supply chain require both—code signing plus tool approval policies. Six categories receive explicit implementations; Multi-Agent Attacks are prevented through independent PEP verification (Lemma 7).

\textbf{Empirical Results}: Our evaluation achieves 100\% recall with zero false positives across 174 test cases. By-design elimination provides deterministic guarantees—violations are cryptographically detected, not pattern-matched. Breaking verification requires solving computationally hard problems (L1) rather than crafting adversarial inputs, explaining why attacks bypassing semantic defenses (Atlas, GitHub MCP CVEs) are deterministically blocked by authenticated workflows.

\noindent\textbf{Real-World Attack Validation.} We validate authenticated workflows against two attacks that compromised production systems—OpenAI's Atlas agentic browser and GitHub's Model Context Protocol server. Both attacks succeeded against baseline defenses using semantic guardrails; both are completely blocked by authenticated workflows through independent verification at control surfaces. We then demonstrate deployment feasibility through enterprise integration scenarios.

\noindent\textbf{OpenAI Atlas Browser Attack.} OpenAI's Atlas agentic browser~\cite{atlas-cve} was compromised through prompt injection causing credential exfiltration (OWASP LLM01, LLM06, LLM02).

We emulate this attack: a legitimate workflow reads financial reports; malicious input embeds hidden instructions to leak credentials. The attack succeeds against semantic guardrails which cannot distinguish adversarial prompts from legitimate document content.

Policy enforcement blocks this cascade. Filesystem policies deny credential paths; network policies restrict external endpoints. PEP verification (Lemma 4) rejects unauthorized operations before execution regardless of LLM reasoning. We tested allowlist, denylist, combinations across framework variants. All blocked the attack with zero false positives.

\noindent\textbf{GitHub MCP Prompt Injection.} GitHub's Model Context Protocol server~\cite{github-mcp-cve} (558,846+ downloads) was compromised through malicious prompts in GitHub issues invoking MCP filesystem tools for credential exfiltration (OWASP LLM01, LLM03, LLM02).

Multiple barriers block this cascade. First, tool authorization requires cryptographic signatures from approved registries—malicious tools rejected at discovery (Lemma 1). Second, filesystem policies block credential access (Lemma 4). Third, cross-framework composition ensures filesystem PEP verifies operations independently (Lemma 5). The attack is blocked through independent verification at multiple surfaces (Lemma 7).

Both attacks demonstrate that application-layer defenses alone are insufficient—production systems were compromised despite semantic guardrails. Authenticated workflows achieve complete protection through defense in depth: multiple independent cryptographic barriers must all be defeated simultaneously. All test scenarios maintained 100\% recall with zero false positives, validating that formal guarantees (Section~\ref{sec:formal-analysis}) translate to production deployments. Section~\ref{sec:related-work} positions these results against existing defenses.

\section{Performance}
\label{sec:performance}

We validate practical overhead through microbenchmarks on commodity hardware (8-core, 16GB RAM) using ECDSA-256 signatures and SHA-256 hashing. \textbf{Cryptographic operations} typically under $\sim$0.2ms overhead (ECDSA signature generation $\sim$102$\mu$s, verification $\sim$89$\mu$s, hash chain updates $<$15$\mu$s per operation). \textbf{Policy operations}: retrieval latency depends on backend: in-memory ($\sim$5-50$\mu$s), filesystem ($\sim$30-200$\mu$s), Redis ($\sim$50-300$\mu$s); policy intersection depends on composition---with typical 3-5 deep hierarchy remains under 100$\mu$s. \textbf{Custom verifiers}: deterministic checks (PII detection, path validation) add 30-400$\mu$s; LLM-based verifiers (prompt safety) add 150-500ms, dominating when enabled but providing semantic analysis orthogonal to cryptographic guarantees. \textbf{End-to-end impact}: For network-bound operations (LLM inference, remote APIs), cryptographic overhead ($\sim$0.2ms) is negligible versus network latency (50-500ms). For local tool invocations, total PEP verification overhead remains under 1ms without LLM verifiers. Audit logging occurs asynchronously via buffered I/O.

A companion paper provides comprehensive system evaluation including production workload characterization, framework integration costs across all nine frameworks, large-scale deployment experience, and detailed performance analysis with production-scale benchmarks.

\section{Related Work}
\label{sec:related-work}

Recent surveys~\cite{acm-ai-agents-survey2024,agentic-security-survey2025,gong-prompt-injection2024,liu-protocol-exploits2025} identify four critical gaps in agentic AI security: unpredictable multi-step workflows, cross-framework compositions, variable environments, and untrusted entity interactions. These surveys diagnose fragmentation but lack systems solutions. Prior work addresses either application semantics or infrastructure primitives; we provide a trust layer positioned between both—protocol-level primitives protecting four fundamental control surfaces (prompts, tools, data, context) where all agent-resource interactions occur.

\textbf{Application-Layer Defenses Fail at Composition.} Framework-specific security operates in isolation. LangChain provides schemas and validation~\cite{langchain2023}; MCP implements authentication and permissions~\cite{mcp2024}; OpenAI enforces rate limits and moderation~\cite{openai-assistants2024}. Recent vulnerabilities demonstrate composition gaps: CVE-2024-8309 (LangChain prompt injection→SQL, CVSS 9.0)~\cite{cve-2024-8309}, CVE-2024-36480 (RCE via unsanitized eval())~\cite{cve-2024-36480}, CVE-2025-6514 (MCP supply chain, 558K+ downloads)~\cite{github-mcp-cve}, Atlas browser attack (October 2025)~\cite{atlas-cve,openai-ciso-atlas}. No cryptographic binding exists across framework boundaries.

Recent academic work addresses prompt injection defenses~\cite{perez2022ignore,liu2023prompt,jain2023baseline}, tool use safety~\cite{ruan2023identifying}, and RAG retrieval poisoning~\cite{zou2024poisonedrag,cheng2024parden}—malicious instructions embedded in documents that compromise reasoning. Semantic defenses (guardrails~\cite{nemoguardrails2023,rebuff2023,llamaguard2023}) achieve 60-80\% detection but remain bypassable~\cite{wei2023jailbroken,zou2023universal}. We provide deterministic guarantees (100\% recall, zero false positives) through defense in depth: by-design elimination renders attacks cryptographically infeasible; by-policy enforcement blocks violations. The approaches are complementary—semantic defenses reduce attack surface; cryptographic enforcement provides non-bypassable verification.

\textbf{Infrastructure Primitives Lack Workflow Semantics.} Infrastructure primitives provide strong guarantees but lack workflow semantics. Identity systems (SPIFFE~\cite{spiffe2023}, AWS IAM~\cite{aws-iam2023}) and policy engines (OPA~\cite{opa2023}, Cedar~\cite{cedar2023}) compose policies through intersection but cannot express temporal dependencies—``operation B requires proof that A completed.'' Cedar composes policies via intersection but lacks: (1) attestation-based workflow dependencies, (2) dynamic principals resolved at runtime, (3) cryptographic binding across four control surfaces. Without cryptographic proof of prerequisite operations, policies must trust application-reported state that attackers can forge.

MAPL addresses this through attestations—unforgeable cryptographic proofs that operations completed, enabling provably correct multi-step workflows. Unlike platform attestation (TPM~\cite{tcg2016}, SGX~\cite{costan2016intel}) proving code integrity through measurements, MAPL attestations prove \textit{workflow completion}—cryptographic proofs that operations executed with specific results, enabling temporal dependencies impossible with measurement-based attestation. We extend authenticated system calls~\cite{rajagopalan2006} and inline reference monitors~\cite{erlingsson2004,schneider2000} from kernel boundaries to four control surfaces, adding stateful verification via authenticated context for workflow dependencies and pluggable verifiers for domain-specific checks (Appendix~\ref{appendix:verifiers}).

\textbf{Orthogonal Approaches.} AI safety~\cite{constitutional_ai2022,ouyang2022training,ganguli2022redteaming} operates at training-time; we provide runtime enforcement. Confidential computing~\cite{costan2016intel,gentry2009fully} protects data confidentiality; we address workflow integrity. These approaches are complementary.

\section{Conclusion}
\label{sec:conclusion}

We presented authenticated workflows—the first complete trust layer for enterprise agentic AI. Our key architectural insight: positioning at the protocol level—between application-layer defenses that fail at composition and infrastructure primitives that lack workflow semantics—enables uniform protection across heterogeneous frameworks while maintaining workflow context. All agent-resource interactions cross four fundamental control surfaces (prompts, tools, data, context); protecting these surfaces uniformly achieves compositional security impossible at either adjacent layer alone.

Three contributions realize this vision: MAPL provides cryptographic workflow dependencies via attestations, enabling sequential constraints without trusting application-reported state; embedded independent PEPs realize defense in depth where multiple cryptographic barriers must be defeated simultaneously; integration across nine frameworks through thin adapters validates surface completeness. Formal proofs (Lemmas 1-7, Theorems 1-3) establish security properties; empirical validation demonstrates 100\% recall, zero false positives across 174 test cases, and protection against two production CVEs that compromised baseline systems. This addresses the five challenges posed in Section~\ref{sec:introduction}: cryptographic integrity across frameworks, policy enforcement at scale, privilege non-escalation through composition, context integrity via hash chains, and complete accountability through non-repudiation. As agentic AI transitions to production, authenticated workflows provide the security substrate enabling safe deployment where composition gaps currently block adoption.

\section*{Acknowledgments}

We thank the anonymous reviewers for their valuable feedback. This work was supported in part by [REDACTED FOR BLIND REVIEW].

{\scriptsize
\bibliographystyle{IEEEtran}
\bibliography{references}

@misc{rebuff2023,
  title={Rebuff: Prompt Injection Detection},
  author={Rebuff.ai},
  year={2023},
  howpublished={\url{https://github.com/protectai/rebuff}}
}

@misc{opa2023,
  title={Open Policy Agent: Policy-based Control},
  author={{CNCF}},
  year={2023},
  howpublished={\url{https://www.openpolicyagent.org/}}
}

@misc{cedar2023,
  title={Cedar: Policy Language for Authorization},
  author={Amazon Web Services},
  year={2023},
  howpublished={\url{https://www.cedarpolicy.com/}}
}

@misc{langchain2023,
  title={LangChain: Building Applications with LLMs},
  author={LangChain Inc.},
  year={2023},
  howpublished={\url{https://www.langchain.com/}}
}

@article{constitutional_ai2022,
  title={Constitutional {AI}: Harmlessness from {AI} Feedback},
  author={Bai, Yuntao and Kadavath, Saurav and Kundu, Sandipan and Askell, Amanda and Kernion, Jackson and Jones, Andy and Chen, Anna and Goldie, Anna and Mirhoseini, Azalia and McKinnon, Cameron and others},
  journal={arXiv preprint arXiv:2212.08073},
  year={2022}
}

@article{ouyang2022training,
  title={Training Language Models to Follow Instructions with Human Feedback},
  author={Ouyang, Long and Wu, Jeffrey and Jiang, Xu and Almeida, Diogo and Wainwright, Carroll and Mishkin, Pamela and Zhang, Chong and Agarwal, Sandhini and Slama, Katarina and Ray, Alex and others},
  journal={Advances in Neural Information Processing Systems},
  volume={35},
  pages={27730--27744},
  year={2022}
}

@article{ganguli2022redteaming,
  title={Red Teaming Language Models to Reduce Harms: Methods, Scaling Behaviors, and Lessons Learned},
  author={Ganguli, Deep and Lovitt, Liane and Kernion, Jackson and Askell, Amanda and Bai, Yuntao and Kadavath, Saurav and Mann, Ben and Perez, Ethan and Schiefer, Nicholas and Ndousse, Kamal and others},
  journal={arXiv preprint arXiv:2209.07858},
  year={2022}
}

@misc{spiffe2023,
  title={{SPIFFE}: Secure Production Identity Framework for Everyone},
  author={{CNCF}},
  year={2023},
  howpublished={\url{https://spiffe.io/}}
}

@article{perez2022ignore,
  title={Ignore Previous Prompt: Attack Techniques For Language Models},
  author={Perez, Fábio and Ribeiro, Ian},
  journal={arXiv preprint arXiv:2211.09527},
  year={2022}
}

@article{wei2023jailbroken,
  title={Jailbroken: How Does {LLM} Safety Training Fail?},
  author={Wei, Alexander and Haghtalab, Nika and Steinhardt, Jacob},
  journal={Advances in Neural Information Processing Systems},
  volume={36},
  year={2023}
}

@article{zou2023universal,
  title={Universal and Transferable Adversarial Attacks on Aligned Language Models},
  author={Zou, Andy and Wang, Zifan and Kolter, J. Zico and Fredrikson, Matt},
  journal={arXiv preprint arXiv:2307.15043},
  year={2023}
}

@misc{nemoguardrails2023,
  title={{NeMo Guardrails}: Programmable Guardrails for {LLM} Applications},
  author={NVIDIA},
  year={2023},
  howpublished={\url{https://github.com/NVIDIA/NeMo-Guardrails}}
}

@article{camel2023,
  title={{CAMEL}: Communicative Agents for Mind Exploration of Large Scale Language Model Society},
  author={Li, Guohao and Hammoud, Hasan Abed Al Kader and Itani, Hani and Khizbullin, Dmitrii and Ghanem, Bernard},
  journal={arXiv preprint arXiv:2303.17760},
  year={2023}
}

@inproceedings{rajagopalan2025prompts,
  title={Protecting Context and Prompts: Deterministic Security for Non-Deterministic {AI}},
  author={Rajagopalan, Mohan and Rao, Vinay},
  organization={MACAW Security, Inc. and ROOST.tools},
  booktitle={Proceedings of the 1st International Workshop on AI Security and Evaluation (IASEAI'26), co-located with IEEE Symposium on Security and Privacy},
  year={2026},
  note={To appear}
}

@article{rajagopalan2006,
  title={Efficient and Flexible Security Support for Distributed Component-Based Systems},
  author={Rajagopalan, Mohan and Hiltunen, Matti A. and Jim, Trevor and Schlichting, Richard D.},
  journal={IEEE Transactions on Dependable and Secure Computing},
  volume={3},
  number={4},
  pages={352--369},
  year={2006},
  publisher={IEEE},
  doi={10.1109/TDSC.2006.41}
}

@article{erlingsson2004,
  title={The Inlined Reference Monitor Approach to Security Policy Enforcement},
  author={Erlingsson, {\'U}lfar and Schneider, Fred B.},
  journal={ACM Transactions on Programming Languages and Systems},
  volume={28},
  number={1},
  pages={2--44},
  year={2004}
}

@article{costan2016intel,
  title={Intel {SGX} Explained},
  author={Costan, Victor and Devadas, Srinivas},
  journal={IACR Cryptology ePrint Archive},
  volume={2016},
  pages={086},
  year={2016}
}

@phdthesis{gentry2009fully,
  title={A Fully Homomorphic Encryption Scheme},
  author={Gentry, Craig},
  year={2009},
  school={Stanford University}
}

@misc{openai-assistants2024,
  title={{OpenAI Assistants API}},
  author={{OpenAI}},
  year={2024},
  howpublished={\url{https://platform.openai.com/docs/assistants/}}
}

@misc{mcp2024,
  title={Model Context Protocol ({MCP})},
  author={{Anthropic}},
  year={2024},
  howpublished={\url{https://modelcontextprotocol.io/}}
}

@misc{llamaguard2023,
  title={{Llama Guard}: {LLM}-based Input-Output Safeguard for Human-AI Conversations},
  author={Inan, Hakan and Upasani, Kartikeya and Chi, Jianfeng and Rungta, Rashi and Iyer, Krithika and Mao, Yuning and Tontchev, Michael and Hu, Qing and Fuller, Brian and Testuggine, Davide and Khabsa, Madian},
  journal={arXiv preprint arXiv:2312.06674},
  year={2023}
}

@misc{cve-2024-8309,
  title={{CVE-2024-8309}: Prompt Injection in {LangChain}'s {GraphCypherQAChain} Leads to Full Database Compromise},
  author={{Huntr Dev}},
  year={2024},
  howpublished={\url{https://nvd.nist.gov/vuln/detail/CVE-2024-8309}}
}

@misc{cve-2024-36480,
  title={{CVE-2024-36480}: Remote Code Execution in {LangChain}},
  author={{NVD}},
  year={2024},
  howpublished={\url{https://nvd.nist.gov/vuln/detail/CVE-2024-36480}},
  note={CVSS 9.0 - Critical}
}

@misc{github-mcp-cve,
  title={{CVE-2025-6514}: Prompt Injection in {GitHub} Model Context Protocol Server},
  author={{Invariant Labs}},
  year={2025},
  month={May},
  howpublished={\url{https://www.invariant.sh/blog/cve-2025-6514}},
  note={Disclosed May 26, 2025; 558,846+ downloads}
}

@misc{atlas-cve,
  title={Prompt Injection Vulnerability in {OpenAI Atlas} Browser},
  author={{Security Researchers}},
  year={2025},
  month={October},
  howpublished={Security Advisory},
  note={Disclosed October 21, 2025}
}

@misc{openai-ciso-atlas,
  title={Statement on Prompt Injection Challenges from {OpenAI CISO}},
  author={{OpenAI}},
  year={2025},
  month={October},
  howpublished={Official Statement}
}

@article{acm-ai-agents-survey2024,
  title={{AI} Agents Under Threat: A Survey of Key Security Challenges and Future Pathways},
  author={Wang, Yanjie and Chen, Xin and Li, Jiahao and Zhang, Wei},
  journal={ACM Computing Surveys},
  year={2024},
  publisher={ACM},
  note={arXiv:2406.02630}
}

@article{agentic-security-survey2025,
  title={A Survey on Agentic Security: Applications, Threats and Defenses},
  author={Zhou, Yifan and Liu, Tianshi and Wang, Jian},
  journal={arXiv preprint arXiv:2510.06445},
  year={2025}
}

@inproceedings{gong-prompt-injection2024,
  title={Formalizing and Benchmarking Prompt Injection Attacks and Defenses},
  author={Gong, Neil and Liu, Yupei and Zhang, Yuancheng},
  booktitle={Proceedings of the 33rd USENIX Security Symposium},
  year={2024},
  organization={USENIX}
}

@article{liu-protocol-exploits2025,
  title={From Prompt Injections to Protocol Exploits: Threats in {LLM}-Powered {AI} Agents Workflows},
  author={Liu, Yang and Chen, Xiang and Wang, Zhiwei},
  journal={arXiv preprint arXiv:2506.23260},
  year={2025},
  month={June}
}

@misc{owasp-llm-2025,
  title={{OWASP} Top 10 for {LLM} Applications 2025},
  author={{OWASP Foundation}},
  year={2025},
  howpublished={\url{https://genai.owasp.org/}}
}

@article{schneider2000,
  title={Enforceable Security Policies},
  author={Schneider, Fred B.},
  journal={ACM Transactions on Information and System Security},
  volume={3},
  number={1},
  pages={30--50},
  year={2000},
  publisher={ACM}
}

@inproceedings{liu2023prompt,
  title={Prompt Injection Attack Against {LLM}-Integrated Applications},
  author={Liu, Yi and Deng, Gelei and Xu, Zhengzi and Li, Yuekang and Zheng, Yaowen and Zhang, Ying and Zhao, Lida and Zhang, Tianwei and Liu, Yang},
  booktitle={Proceedings of the 2023 ACM SIGSAC Conference on Computer and Communications Security (CCS)},
  pages={1-15},
  year={2023},
  publisher={ACM}
}

@article{jain2023baseline,
  title={Baseline Defenses for Adversarial Attacks Against Aligned Language Models},
  author={Jain, Neel and Schwarzschild, Avi and Wen, Yuxin and Somepalli, Gowthami and Kirchenbauer, John and Chiang, Ping-yeh and Goldblum, Micah and Saha, Aniruddha and Geiping, Jonas and Goldstein, Tom},
  journal={arXiv preprint arXiv:2309.00614},
  year={2023}
}

@article{ruan2023identifying,
  title={Identifying the Risks of {LM} Agents with an {LM}-Emulated Sandbox},
  author={Ruan, Yangjun and Dong, Honghua and Wang, Andrew and Pitis, Silviu and Zhou, Yongchao and Ba, Jimmy and Dubois, Yann and Maddison, Chris J and Hashimoto, Tatsunori},
  journal={arXiv preprint arXiv:2309.15817},
  year={2023}
}

@inproceedings{zou2024poisonedrag,
  title={{PoisonedRAG}: Knowledge Poisoning Attacks to Retrieval-Augmented Generation of Large Language Models},
  author={Zou, Wei and Geng, Runpeng and Wang, Binghui and Jia, Jinyuan},
  booktitle={arXiv preprint arXiv:2402.07867},
  year={2024}
}

@article{cheng2024parden,
  title={{PARDEN}: Toward Detecting Backdoor Attacks in Retrieval-Augmented Generation},
  author={Cheng, Zhenting and Li, Guanghan and Liu, Jiaxin and Zhang, Qingfu and Shao, Jining},
  journal={arXiv preprint arXiv:2409.01573},
  year={2024}
}

@misc{aws-iam2023,
  title={{AWS} Identity and Access Management: Policy Composition and Authorization},
  author={{Amazon Web Services}},
  year={2023},
  howpublished={\url{https://aws.amazon.com/iam/}}
}

@techreport{tcg2016,
  title={{TPM} 2.0 Library Specification},
  author={{Trusted Computing Group}},
  year={2016},
  institution={Trusted Computing Group},
  note={Specification Family "2.0", Level 00 Revision 01.38}
}
}

\appendix
\section*{Appendix A: MAPL Policy Examples}
\label{appendix:policy-examples}

{\small
\noindent\begin{minipage}[t]{0.48\columnwidth}
\textbf{Base Organizational Policy}
\begin{lstlisting}[basicstyle=\ttfamily\scriptsize]
{
  "policy_id": "acme:base",
  "resources": ["tool:*", "llm:*"],
  "denied_resources": ["*credential*"],
  "constraints": {
    "attestations": ["user_authenticated"]
  }
}
\end{lstlisting}
Baseline with credential blocks.
\end{minipage}
\hfill
\begin{minipage}[t]{0.48\columnwidth}
\textbf{Department Policy (Finance)}
\begin{lstlisting}[basicstyle=\ttfamily\scriptsize]
{
  "policy_id": "acme:finance",
  "extends": "acme:base",
  "resources": ["data:finance:**"],
  "constraints": {
    "attestations": ["mfa_verified"]
  }
}
\end{lstlisting}
Inherits base, adds MFA.
\end{minipage}

\vspace{0.5em}
\noindent\begin{minipage}[t]{0.48\columnwidth}
\textbf{Attestation Chaining}
\begin{lstlisting}[basicstyle=\ttfamily\scriptsize]
// Step 1: Analysis
{"policy_id": "fin:analyze",
 "resources": ["tool:analyze"],
 "constraints": {"attestations":
   ["user_authenticated"]}}
// -> Produces: analysis_done

// Step 2: Requires Step 1
{"policy_id": "fin:report",
 "resources": ["tool:report"],
 "constraints": {"attestations":
   ["analysis_done"]}}
// -> Produces: report_done

// Step 3: Requires Step 2
{"policy_id": "fin:send",
 "resources": ["tool:email"],
 "constraints": {"attestations":
   ["report_done"]}}
\end{lstlisting}
Cryptographic proof prevents workflow hijacking.
\end{minipage}
\hfill
\begin{minipage}[t]{0.48\columnwidth}
\textbf{Resource Policy (Database)}
\begin{lstlisting}[basicstyle=\ttfamily\scriptsize]
{
  "policy_id": "database:customer_db",
  "constraints": {
    "denied_parameters": {
      "query": ["DROP", "DELETE"]
    }
  }
}
\end{lstlisting}
Dual-perspective constraints.
\end{minipage}

\vspace{0.5em}
\noindent\begin{minipage}[t]{0.48\columnwidth}
\textbf{Delegation Policy}
\begin{lstlisting}[basicstyle=\ttfamily\scriptsize]
{
  "policy_id": "agent:sub_agent",
  "extends": "acme:finance",
  "resources": ["data:finance:reports:*"],
  "constraints": {
    "parameters": {"date_range":
      ["2024-Q1", "2024-Q2"]}
  }
}
\end{lstlisting}
Narrowed delegation scope.
\end{minipage}
\hfill
\begin{minipage}[t]{0.48\columnwidth}
\textbf{Emergency Access}
\begin{lstlisting}[basicstyle=\ttfamily\scriptsize]
{
  "policy_id": "emergency:incident",
  "extends": "acme:base",
  "validity": {
    "not_after": "2024-10-25T16:00:00Z"
  },
  "constraints": {"attestations":
    ["ciso_approved"]}
}
\end{lstlisting}
Time-bounded (2-hour window).
\end{minipage}
}

\section*{Appendix B: Custom Verifier Implementation}
\label{appendix:verifiers}

\subsection*{Verification Pipeline Architecture}

{\tiny
The PEP verification pipeline composes five stages:

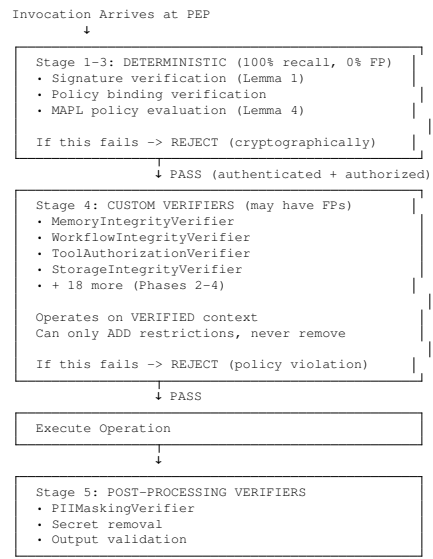
\begin{verbatim}
Invocation Arrives at PEP
         ↓
┌──────────────────────────────────────────────────┐
│  Stage 1-3: DETERMINISTIC (100% recall, 0% FP)  │
│  • Signature verification (Lemma 1)             │
│  • Policy binding verification                   │
│  • MAPL policy evaluation (Lemma 4)             │
│                                                   │
│  If this fails -> REJECT (cryptographically)    │
└─────────────────┬────────────────────────────────┘
                  ↓ PASS (authenticated + authorized)
┌──────────────────────────────────────────────────┐
│  Stage 4: CUSTOM VERIFIERS (may have FPs)       │
│  • MemoryIntegrityVerifier                       │
│  • WorkflowIntegrityVerifier                     │
│  • ToolAuthorizationVerifier                     │
│  • StorageIntegrityVerifier                      │
│  • + 18 more (Phases 2-4)                       │
│                                                   │
│  Operates on VERIFIED context                    │
│  Can only ADD restrictions, never remove         │
│                                                   │
│  If this fails -> REJECT (policy violation)     │
└─────────────────┬────────────────────────────────┘
                  ↓ PASS
┌──────────────────────────────────────────────────┐
│  Execute Operation                               │
└─────────────────┬────────────────────────────────┘
                  ↓
┌──────────────────────────────────────────────────┐
│  Stage 5: POST-PROCESSING VERIFIERS              │
│  • PIIMaskingVerifier                            │
│  • Secret removal                                │
│  • Output validation                             │
└──────────────────────────────────────────────────┘
\end{verbatim}

Even if custom verifiers are compromised, Stages 1-3 enforce cryptographic integrity and policy constraints.
}

\subsection*{Production Verifiers}

Four production verifiers address OWASP Top 10 threats through domain-specific checks on cryptographically verified context:

{\tiny
\noindent\begin{minipage}[t]{0.48\columnwidth}
\textbf{MemoryIntegrityVerifier}\\
\textit{Prevents memory poisoning (OWASP LLM01)}\\
• Rate limiting (10 updates/min)\\
• Protected fields (goals, system\_prompt)\\
• Pattern detection (suspicious updates)\\
• Cryptographic integrity checks
\end{minipage}
\hfill
\begin{minipage}[t]{0.48\columnwidth}
\textbf{WorkflowIntegrityVerifier}\\
\textit{Prevents workflow hijacking (OWASP LLM04)}\\
• Sequence enforcement (prerequisites)\\
• State tracking (prevent re-execution)\\
• Attestation chains (crypto proof)\\
• Duration limits (max 3600s)
\end{minipage}

\vspace{0.5em}

\noindent\begin{minipage}[t]{0.48\columnwidth}
\textbf{ToolAuthorizationVerifier}\\
\textit{Prevents tool misuse (OWASP LLM06)}\\
• RBAC (role-to-tool mapping)\\
• Dangerous patterns (write→execute)\\
• Usage rate limiting (30 tools/min)\\
• Privilege escalation prevention
\end{minipage}
\hfill
\begin{minipage}[t]{0.48\columnwidth}
\textbf{StorageIntegrityVerifier}\\
\textit{Prevents data exfiltration (OWASP LLM02)}\\
• Path traversal prevention\\
• Data classification (secret/sensitive)\\
• Encryption enforcement\\
• Exfiltration detection (bulk reads)
\end{minipage}
}

\section*{Appendix C: Framework Integration Details}
\label{appendix:framework-details}

{\scriptsize
Detailed mappings showing how framework constructs map to authenticated workflow primitives across nine frameworks.

\subsection*{A2A (Agent-to-Agent Protocol)}

A2A defines standards for agent discovery, trust establishment, and communication. MACAW provides the cryptographic infrastructure A2A requires:

{\tiny
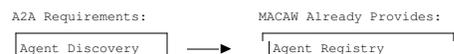
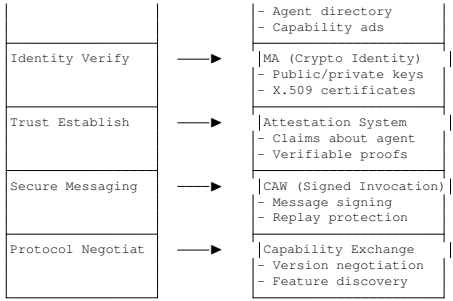
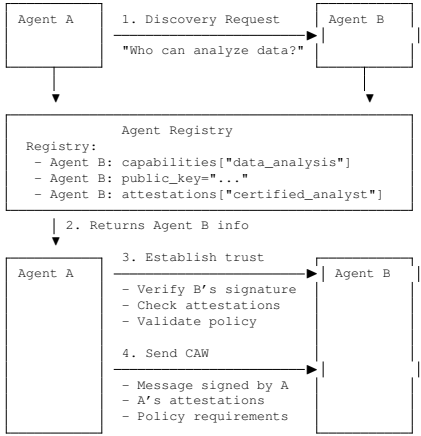
\begin{verbatim}
A2A Requirements:              MACAW Already Provides:
┌──────────────────┐           ┌───────────────────────┐
│Agent Discovery   │  ────►    │Agent Registry         │
│                  │           │- Agent directory      │
│                  │           │- Capability ads       │
├──────────────────┤           ├───────────────────────┤
│Identity Verify   │  ────►    │MA (Crypto Identity)   │
│                  │           │- Public/private keys  │
│                  │           │- X.509 certificates   │
├──────────────────┤           ├───────────────────────┤
│Trust Establish   │  ────►    │Attestation System     │
│                  │           │- Claims about agent   │
│                  │           │- Verifiable proofs    │
├──────────────────┤           ├───────────────────────┤
│Secure Messaging  │  ────►    │CAW (Signed Invocation)│
│                  │           │- Message signing      │
│                  │           │- Replay protection    │
├──────────────────┤           ├───────────────────────┤
│Protocol Negotiat │  ────►    │Capability Exchange    │
│                  │           │- Version negotiation  │
│                  │           │- Feature discovery    │
└──────────────────┘           └───────────────────────┘

Complete A2A Flow:
┌───────────┐                          ┌───────────┐
│ Agent A   │  1. Discovery Request    │ Agent B   │
│           │ ────────────────────────►│           │
│           │  "Who can analyze data?" │           │
└─────┬─────┘                          └─────┬─────┘
      │                                      │
      ▼                                      ▼
┌──────────────────────────────────────────────────┐
│              Agent Registry                      │
│  Registry:                                       │
│   - Agent B: capabilities["data_analysis"]       │
│   - Agent B: public_key="..."                    │
│   - Agent B: attestations["certified_analyst"]   │
└──────────────────────────────────────────────────┘
      │ 2. Returns Agent B info
      ▼
┌───────────┐  3. Establish trust      ┌───────────┐
│ Agent A   │ ────────────────────────►│ Agent B   │
│           │  - Verify B's signature  │           │
│           │  - Check attestations    │           │
│           │  - Validate policy       │           │
│           │                          │           │
│           │  4. Send CAW             │           │
│           │ ────────────────────────►│           │
│           │  - Message signed by A   │           │
│           │  - A's attestations      │           │
│           │  - Policy requirements   │           │
└───────────┘                          └───────────┘
\end{verbatim}}

\textbf{Security Challenge}: Agents must verify identity and capabilities before delegating authority. Attestations provide cryptographic claims that enable trust decisions without centralized authorities.

\subsection*{MCP (Model Context Protocol)}

MCP servers expose three capability types: tools (executable functions), resources (data access with read/write/watch operations), and prompts (retrievable templates). The MACAW mapping:

{\tiny
\begin{verbatim}
MCP Core Concepts:
┌─────────────────────────────────────────────┐
│              MCP Server                     │
├──────────┬──────────────┬───────────────────┤
│  Tools   │  Resources   │     Prompts       │
│execute() │ read/write() │ getPrompt()       │
└──────────┴──────────────┴───────────────────┘

MACAW Mapping:
┌─────────────────────────────────────────────┐
│         macawAgent (MCP Server)             │
├──────────┬──────────────┬───────────────────┤
│  Tool    │ Tool         │ Authenticated     │
│  (execute│ (Resource)   │ Prompt            │
│   → CAW) │ (read → CAW) │ (signed template) │
└──────────┴──────────────┴───────────────────┘

Resource Flow:
Client: "Read file.txt" → SecureMCP: invoke_tool()
  → PEP: signature valid? → Policy: allow read?
  → Execute if allowed
\end{verbatim}}

\textbf{Security Challenge}: Resources represent read-only data access, while tools can modify state. The mapping enforces read-only constraints through policy, preventing privilege escalation from data retrieval to data modification.

\subsection*{OpenAI and Claude (LLM Interfaces)}

LLM providers expose inference APIs with function calling. Two deployment variants exist based on trust assumptions:

{\tiny
\begin{verbatim}
Two-sided Wrapping: Both LLM API and Client as macawAgents
┌─────────────────────────────────────────────────┐
│            Application Code                     │
└───────────────┬─────────────────────────────────┘
                ▼
┌─────────────────────────────────────────────────┐
│          SecureOpenAI/SecureClaude              │
├────────────────────┬────────────────────────────┤
│  LLM Wrapper       │   Tool Wrapper             │
│                    │                            │
│ chat.complete()    │  search_tool()             │
│      ↓             │       ↓                    │
│  INTERCEPT         │   INTERCEPT                │
│      ↓             │       ↓                    │
│  Sign as macawAgent│   Sign invocation          │
│      ↓             │       ↓                    │
│  PEP verifies:     │   PEP verifies:            │
│  - Signature       │   - Signature              │
│  - Provider policy │   - App policy             │
│      ↓             │       ↓                    │
│  Execute LLM call  │   Execute tool             │
└────────────────────┴────────────────────────────┘

One-sided Wrapping: Only Client Tools as macawAgents
┌─────────────────────────────────────────────────┐
│            Application Code                     │
└───────────────┬─────────────────────────────────┘
                ▼
┌─────────────────────────────────────────────────┐
│          SecureOpenAI/SecureClaude              │
├────────────────────┬────────────────────────────┤
│  LLM Passthrough   │   Tool Wrapper             │
│                    │                            │
│ chat.complete()    │  search_tool()             │
│      ↓             │       ↓                    │
│  Direct to API     │   INTERCEPT                │
│  (no interception) │       ↓                    │
│                    │   Sign invocation          │
│                    │       ↓                    │
│                    │   PEP verifies:            │
│                    │   - Signature              │
│                    │   - Policy                 │
│                    │       ↓                    │
│                    │   Execute tool             │
└────────────────────┴────────────────────────────┘
\end{verbatim}}

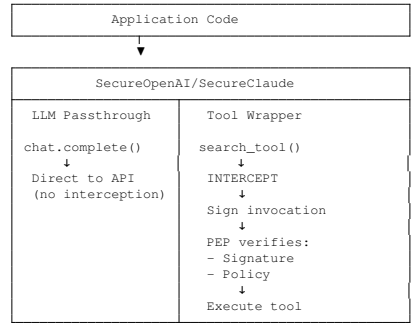
\textbf{Security Challenge}: LLMs are untrusted for authorization decisions. Even with prompt engineering, adversarial prompts can cause unauthorized function calls. PEPs verify at function execution boundaries regardless of LLM decisions.

\subsection*{LangChain (Multi-Agent Orchestration)}

LangChain coordinates multi-step workflows across tools and LLMs. Multi-level integration addresses four layers:

{\tiny
\begin{verbatim}
LangChain Architecture:
┌─────────────────────────────────────────────────┐
│         LangChain Orchestrator                  │
│              (macawAgent)                       │
├──────────┬─────────┬──────────┬─────────────────┤
│ Agents   │ Tools   │  Chains  │    Memory       │
│          │         │          │                 │
│Coordinate│Individual│Sequential│Persistent state │
│workflow  │operations│compose   │across turns     │
│          │         │          │                 │
│Agent →   │Tool →   │Chain →   │Memory →         │
│macawAgent│ Tool    │ Policy   │AuthenticatedCntx│
│          │ (w/PEP) │ Intersect│ (hash chains)   │
│          │         │          │                 │
│Each agent│Each tool│Effective │Context integrity│
│has own   │verifies │policy =  │verified every   │
│identity  │CAW      │∩ of all  │state transition │
│and policy│before   │tool      │                 │
│          │execution│policies  │Prevents context │
│          │         │          │poisoning attacks│
└──────────┴─────────┴──────────┴─────────────────┘

Workflow Flow:
User: "Analyze documents and send summary"
  ↓
LangChain Agent (macawAgent)
  ↓
Step 1: Read docs
  → Tool invocation → CAW → PEP check → Execute
  ↓
Step 2: Analyze content
  → LLM invocation → CAW → PEP check → Execute
  ↓
Step 3: Send email
  → Tool invocation → CAW → PEP check → Execute

Policy Composition via Intersection:
If tool_1 policy allows {A, B, C}
And tool_2 policy allows {B, C, D}
And chain composes tool_1 → tool_2
Then effective policy = {B, C} (intersection)
Chain composition cannot relax tool policies.
\end{verbatim}}

\textbf{Security Challenge}: Orchestrators expose multiple control points beyond individual tools—chain composition logic, agent reasoning, memory operations. Securing only tools would leave composition and state management unprotected.

\subsection*{CrewAI (Role-Based Multi-Agent)}

CrewAI organizes agents into crews where each member has a role executing tasks collaboratively:

{\tiny
\begin{verbatim}
CrewAI Architecture:
┌─────────────────────────────────────────────────┐
│                   Crew                          │
│              (macawAgent)                       │
├──────────┬─────────┬──────────┬─────────────────┤
│ Members  │  Roles  │  Tasks   │  Collaboration  │
│          │         │          │                 │
│Agents    │Role-    │Work items│Inter-agent      │
│with      │based    │assigned  │communication    │
│specific  │perms    │to agents │                 │
│roles     │         │          │                 │
│          │         │          │                 │
│Member →  │Role →   │Task →    │Message →        │
│macawAgent│Attestat │ CAW      │ CAW with        │
│          │ ion     │          │ role check      │
│          │         │          │                 │
│Writer    │"writer" │Execute   │Writer cannot    │
│cannot    │attested │task only │invoke DB tools  │
│access DB │crypto-  │if role   │reserved for     │
│tools     │graphical│matches   │Researcher role  │
└──────────┴─────────┴──────────┴─────────────────┘

Role-Based Flow:
Crew: Researcher + Writer
  ↓
Task: "Write report on Q4 data"
  ↓
Researcher (macawAgent with "researcher" attestation):
  → Invokes database_query tool
  → PEP checks: signature valid? role="researcher"?
  → Policy: "database_query" requires attestation["researcher"]
  → Allowed: execute query
  ↓
Writer (macawAgent with "writer" attestation):
  → Receives data from Researcher
  → Attempts database_query tool
  → PEP checks: signature valid? role="writer"?
  → Policy: "database_query" requires attestation["researcher"]
  → Denied: Writer lacks required attestation
\end{verbatim}}

\textbf{Security Challenge}: Role-based permissions must be cryptographically enforced. A ``writer'' role should not access databases reserved for ``researchers,'' even if the LLM powering the writer attempts unauthorized invocations. Attestations bind roles cryptographically to agent identities.

\subsection*{AutoGen (Autonomous Code Generation)}

AutoGen enables autonomous code generation and execution. The security challenge is unrestricted code execution:

{\tiny
\begin{verbatim}
AutoGen Architecture:
┌─────────────────────────────────────────────────┐
│         AutoGen Conversation                    │
│              (macawAgent)                       │
├──────────┬─────────┬──────────┬─────────────────┤
│ Agents   │ Code    │  Exec    │   Validation    │
│          │  Gen    │          │                 │
│Conversat │Generate │Execute   │Verify code      │
│ional     │Python   │generated │before execution │
│agents    │code to  │code      │                 │
│          │solve    │          │                 │
│          │tasks    │          │                 │
│          │         │          │                 │
│Agent →   │LLM →    │Exec →    │Verifiers:       │
│macawAgent│ Tool    │ Tool     │- AST analysis   │
│          │         │ (w/PEP)  │- Allowlist check│
│          │         │          │- Sandbox verify │
│          │         │          │                 │
│Each agent│Code gen │Execution │Pre-invocation   │
│signs     │becomes  │request = │verifiers analyze│
│requests  │CAW with │CAW       │code content     │
│          │code     │          │before execution │
└──────────┴─────────┴──────────┴─────────────────┘

Code Execution Flow:
User: "Calculate Fibonacci(10)"
  ↓
AutoGen Agent generates code:
  def fib(n): return n if n<2 else fib(n-1)+fib(n-2)
  ↓
Execution request → CAW
  ↓
PEP invokes pre-execution verifiers:
  1. AST analysis: detect dangerous patterns
     (os.system, subprocess, eval, exec, import sys)
  2. Allowlist check: only math/basic operations
  3. Sandbox verification: no network/file access
  ↓
Verifiers approve → Execute in sandbox → Return result
  ↓
Malicious attempt: "os.system('rm -rf /')"
  ↓
AST analysis detects os.system → DENY
\end{verbatim}}

\textbf{Security Challenge}: Unrestricted code execution enables arbitrary operations including data exfiltration and privilege escalation. Pluggable verifiers analyze code through AST parsing before execution, enforcing allowlists and sandbox constraints.

\subsection*{LlamaIndex (RAG Pipelines)}

LlamaIndex provides RAG (Retrieval Augmented Generation) pipelines for document processing and question answering:

{\tiny
\begin{verbatim}
LlamaIndex Architecture:
┌─────────────────────────────────────────────────┐
│         LlamaIndex Pipeline                     │
│              (macawAgent)                       │
├──────────┬─────────┬──────────┬─────────────────┤
│  Index   │ Query   │ Retrieve │   Generate      │
│          │  Engine │          │                 │
│Document  │Process  │Vector    │LLM generates    │
│indexing  │queries  │search    │answer from      │
│          │         │          │retrieved docs   │
│          │         │          │                 │
│Index →   │Query →  │Retrieve →│Generate →       │
│macawAgent│ Tool    │ Tool     │ Tool            │
│          │ (w/PEP) │ (w/PEP)  │ (w/PEP)         │
│          │         │          │                 │
│Vector DB │Query    │Document  │LLM receives     │
│becomes   │validated│access    │only allowed     │
│macawAgent│against  │validated │documents        │
│          │policy   │against   │                 │
│          │         │policy    │                 │
└──────────┴─────────┴──────────┴─────────────────┘

RAG Flow with Security:
User: "What were Q4 earnings?"
  ↓
Query Engine (Tool with PEP)
  → CAW: query="Q4 earnings"
  → PEP checks:
     - Signature valid?
     - Policy allows query pattern?
     - Denied patterns: *password*, *credentials*
  → Approved: continue
  ↓
Retriever (Tool with PEP)
  → CAW: retrieve docs matching "Q4 earnings"
  → PEP checks:
     - Can access these documents?
     - Allowed files: *public*, *reports*
     - Denied files: *private*, *secret*
  → Approved: retrieve documents
  ↓
Generator (Tool with PEP)
  → CAW: generate answer from retrieved docs
  → PEP checks: output validation
  → Return answer to user
\end{verbatim}}

\textbf{Security Challenge}: RAG pipelines access untrusted document stores. Malicious documents could contain prompt injection or sensitive data. PEPs verify document access patterns and query constraints before retrieval.

\subsection*{Haystack (Document Processing Pipelines)}

Haystack provides document processing pipelines with nodes for retrieval, preprocessing, and generation:

{\tiny
\begin{verbatim}
Haystack Architecture:
┌─────────────────────────────────────────────────┐
│         Haystack Pipeline                       │
│              (macawAgent)                       │
├──────────┬─────────┬──────────┬─────────────────┤
│ Pipeline │  Nodes  │ Retrieve │   Process       │
│          │         │          │                 │
│Sequential│Pipeline │Document  │Text processing  │
│execution │steps    │retrieval │and generation   │
│of nodes  │         │          │                 │
│          │         │          │                 │
│Pipeline →│Node →   │Retrieve →│Process →        │
│macawAgent│ Tool    │ Tool     │ Tool            │
│          │ (w/PEP) │ (w/PEP)  │ (w/PEP)         │
│          │         │          │                 │
│Each node │Each node│Document  │Processing       │
│in chain  │verifies │access    │validated        │
│enforces  │CAW      │controlled│against policy   │
│policy    │         │by policy │                 │
└──────────┴─────────┴──────────┴─────────────────┘

Pipeline Flow:
User: "Process documents in ./reports/"
  ↓
Pipeline (macawAgent)
  ↓
Node 1: FileReader (Tool with PEP)
  → CAW: read("./reports/")
  → PEP checks:
     - Allowed paths: *reports*, *public*
     - Denied paths: *credentials*, *private*
  → Approved: read files
  ↓
Node 2: Preprocessor (Tool with PEP)
  → CAW: clean text, remove PII
  → PEP checks: PII detection verifier
  → Execute with redaction
  ↓
Node 3: Generator (Tool with PEP)
  → CAW: generate summary
  → PEP checks: output validation
  → Return result

Policy Composition Across Nodes:
If Node 1 allows {A, B, C}
And Node 2 allows {B, C, D}
And pipeline chains Node 1 → Node 2
Then effective policy = {B, C} (intersection)
\end{verbatim}}

\textbf{Security Challenge}: Document processing pipelines access file systems and external data sources. Nodes must enforce path constraints and content validation to prevent unauthorized access or data exfiltration.

\section*{Appendix D: AI Agent Threat Taxonomy}

\noindent
{\scriptsize
\setlength{\tabcolsep}{3pt}
\begin{tabular}{|l|p{3cm}|p{12.8cm}|}
\hline
\textbf{ID} & \textbf{Attack} & \textbf{Description \& Impact} \\
\hline
\multicolumn{3}{|l|}{\textbf{Category A: Prompt Injection \& Input Manipulation}} \\
\hline
T-A1 & Hidden Prompt Injection & Embedded via white-on-white text, CSS, HTML comments. Impact: Command execution, credential theft. \\
T-A2 & Semantic Obfuscation & Rephrasing to evade filters, base64 encoding. Impact: Filter bypass. \\
T-A3 & Multi-Turn Poisoning & Gradual injection across turns, fabricated history. Impact: Privilege escalation. \\
T-A4 & Delayed Activation & Conditional triggers (time-bombs). Impact: Evading immediate detection. \\
\hline
\multicolumn{3}{|l|}{\textbf{Category B: Tool Chaining \& Derivation Exploits}} \\
\hline
T-B1 & Privilege Escalation via Chaining & Combining benign operations (search→list→read). Impact: Policy circumvention. \\
T-B2 & Semantic Drift & Progressive deviation from intent. Impact: Compounding interpretation errors. \\
T-B3 & Derivation Depth Exhaustion & Deep recursion chains. Impact: Resource exhaustion, policy drift. \\
\hline
\multicolumn{3}{|l|}{\textbf{Category C: Credential \& Session Attacks}} \\
\hline
T-C1 & Token Extraction & Reading credentials from context. Impact: Session hijacking, cloud access. \\
T-C2 & Credential Exfiltration & Tool abuse sequence (read→email). Impact: Persistent unauthorized access. \\
T-C3 & OAuth Flow Manipulation & Tricking OAuth to attacker apps. Impact: Account takeover. \\
T-C4 & Session Replay & Reusing valid signed prompts. Impact: Stale authentication exploitation. \\
\hline
\multicolumn{3}{|l|}{\textbf{Category D: Data Exfiltration \& Privacy Breaches}} \\
\hline
T-D1 & Conversation History Mining & Extracting PII/secrets from history. Impact: GDPR/HIPAA violations, IP theft. \\
T-D2 & Cross-Principal Leakage & Insufficient context isolation. Impact: Unauthorized cross-user access. \\
T-D3 & Embedding Space Poisoning & Polluting vector DBs/RAG. Impact: Persistent data contamination. \\
T-D4 & Document Harvesting & Systematic file extraction. Impact: Mass data theft. \\
\hline
\multicolumn{3}{|l|}{\textbf{Category E: Malware \& Code Execution}} \\
\hline
T-E1 & AI-Invoked Malware & Downloads/executes malicious code. Impact: Ransomware, backdoors. \\
T-E2 & Code Generation Exploitation & Generates disguised malicious scripts. Impact: Resource hijacking. \\
T-E3 & Persistence Installation & Creating backdoors (.bashrc, cron). Impact: Persistent compromise. \\
T-E4 & Supply Chain via Packages & Installing trojanized dependencies. Impact: Dependency confusion. \\
\hline
\multicolumn{3}{|l|}{\textbf{Category F: Resource Exhaustion \& Cost Attacks}} \\
\hline
T-F1 & Computational Exhaustion & Recursive ops, infinite loops. Impact: DoS, billing inflation. \\
T-F2 & API Rate Limit Exhaustion & Depleting quotas via loops. Impact: Service degradation, \$1000s+ costs. \\
T-F3 & Storage/Bandwidth Saturation & Unbounded downloads/generation. Impact: System failure. \\
\hline
\multicolumn{3}{|l|}{\textbf{Category G: Multi-Agent System Attacks}} \\
\hline
T-G1 & Agent-to-Agent Infection & Poisoned prompts via MCP/A2A. Impact: Cascading compromise. \\
T-G2 & Workflow Hijacking & Crafted inter-agent requests. Impact: Distributed policy bypass. \\
T-G3 & Byzantine Agent Behavior & Malicious agent with valid credentials. Impact: Authorized attacks. \\
T-G4 & Trust Anchor Compromise & Corrupting registry/identity provider. Impact: Complete compromise. \\
\hline
\multicolumn{3}{|l|}{\textbf{Category H: Policy \& Compliance Violations}} \\
\hline
T-H1 & Policy Confusion & Exploiting rule ambiguities. Impact: Authorization gaps. \\
T-H2 & Audit Trail Manipulation & Corrupting logs, false entries. Impact: Loss of forensic evidence. \\
T-H3 & Compliance Bypass & GDPR/HIPAA/SOC2 violations. Impact: Legal liability, 4\% revenue fines. \\
T-H4 & Attestation Forgery & Fake security check proofs. Impact: Workflow security bypass. \\
\hline
\end{tabular}

\vspace{2mm}
\noindent\textit{Comprehensive taxonomy of 25 attack variants across eight categories (T-A through T-H). Section 8 validates coverage.}
}

\end{document}